\documentclass[aps,twocolumn,showpacs,amsmath,amssymb,floatfix,prb,superscriptaddress]{revtex4-1}

\usepackage{graphicx,float}
\usepackage{subfigure}
\usepackage{dcolumn}
\usepackage{bm}
\usepackage{mathrsfs}
\usepackage{txfonts}
\usepackage{CJK}
\usepackage{SIunits}

\usepackage{color}

\allowdisplaybreaks[2]

\newenvironment{SChinese}{%
  \CJKfamily{gbsn}%
 \CJKtilde
  \CJKnospace}{}

\begin{document}

\begin{CJK}{UTF8}{}
\begin{SChinese}

\title{Stationary entanglement in strongly coupled qubits}
\author{Keyu Xia (夏可宇)}
\email{keyu.xia@mpi-hd.mpg.de}
\affiliation{Max-Planck-Institut f\"{u}r Kernphysik, Saupfercheckweg 1, D-69117 Heidelberg, Germany}
\affiliation{Institute for Quantum Studies and Department of Physics and Astronomy, Texas A\&M University, College Station, Texas 77843-4242, USA}

\author{Mihai Macovei}%
\email{Mihai.Macovei@mpi-hd.mpg.de}
\affiliation{Max-Planck-Institut f\"{u}r Kernphysik, Saupfercheckweg 1, D-69117 Heidelberg, Germany}

\author{J\"{o}rg Evers}%
\email{joerg.evers@mpi-hd.mpg.de}
\affiliation{Max-Planck-Institut f\"{u}r Kernphysik, Saupfercheckweg 1, D-69117 Heidelberg, Germany}

\date{\today}

\begin{abstract}
The dynamics of two superconducting flux qubits coupled to each other and to a common bath is discussed. We focus on the case in which the qubit-qubit coupling strength dominates over the respective qubit transition frequencies.  We derive the  master equation including collective effect by modeling the bath as 1D open space in this ultra-strong coupling regime, and find that the coupling greatly modifies both the coherent and the incoherent dynamics of the system, giving rise to qualitatively different properties. 
By analyzing the steady-state and the dynamics governed by the master equation, we show that ground state entanglement and maximum coherence between the two qubits can be induced by the environment alone. By employing in addition a single external driving field, both the  entangled anti-symmetric and symmetric collective states can be populated and preserved with high fidelity. Similarly, entangled states can be prepared using adiabatic passage techniques using two external fields. Our results could find applications in entangling quantum gates  and quantum memories free from the decoherence.
\end{abstract}

\pacs{85.25.Cp, 03.67.Bg, 03.67.Pp, 42.50.Hz}


\maketitle
\end{SChinese}
\end{CJK}

\section{introduction}
Entanglement is the primary resource for quantum computation and information processing~\cite{ChuangNielsen}, and correspondingly, efficient entanglement generation and preservation is a subject of intense research in all physical systems considered as possible realization for quantum protocols. Here, we focus on superconducting qubits~\cite{Nature474p589,Nature453p1031}, which in principle have the particular advantage of great freedom in designing the system properties. They promise an attractive route towards a scalable quantum computer, such that creation and protection of entanglement in superconducting qubits is of particular interest.
Consequently, entanglement generation between superconducting qubits has been suggested and demonstrated recently  by a number of groups \cite{PRA79p052308,PRB76p100505R,PRL93p037003,PRL96p067003,Science300p1548,Nature461p504,PRL94p240502,PRL94p027003,PRB79p024519}. Usually, the creation of entanglement requires precise control of all fields interacting with the qubits. However, in typical experiments, the inevitable decoherence and the imperfection of actual driving fields reduce the achievable fidelity for state preparation and quantum gates.

The decoherence, which arises due to the coupling of a system with its environment, usually tends to destroy the quantum effects and entanglement at the heart of quantum protocols which provide the advantages over corresponding classic algorithms. 
Superconducting qubits are particularly limited in their coherence time due to the many uncontrollable degrees of freedom in their solid-state environment.
Several proposals have been made to eliminate noise-induced decoherence \cite{NaturePhys5p633,PRA78p042307,PRL89p277901,PRL91p070402,NaturePhys5p48,PRL100p113601,PRB81p014505,PRA79p052308,PRB66p054527,PRB76p100505R,PRB79p024519,PRL89p207902}. In particular, by careful engineering of the  local environment coupling to a system, entangled pair states have been prepared using a dissipative process \cite{NaturePhys5p633,PRA78p042307}. However it is difficult in experiments to isolate well the local environments coupled to the individual qubits  from each other. Moreover, for a given setup, the steady state which can be prepared is unique. Alternatively, entanglement also has been created between two qubits interacting with a common heat bath~\cite{PRL89p277901,PRL91p070402}.  

\begin{figure}[b]
 \centering
 \includegraphics[width=0.9\columnwidth]{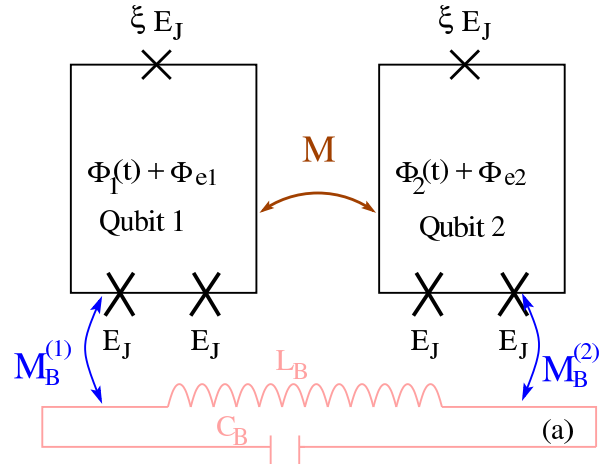}
 \caption{\label{fig:system}(Color online) The considered model system. Two superconducting flux qubits are coupled through their mutual effective inductance $M$ and coupled to a common reservoir modeled as an LC circuit via the mutual inductance $M_B$. We consider the case of strong coupling between the qubits, in which the always-on coupling strength between the two qubits exceeds their respective transition frequencies.}
\end{figure}

Next to the creation, also the preservation of entangled states is a long-standing challenge.
As a possible solution, it has been pointed out that ground state entanglement is intrinsically immune to the decoherence. It has been discussed in complex system \cite{PRA80p043825,PRA80p063606}, but without addressing the inevitable decoherence.
An alternative is the well studied steady state entanglement \cite{PRA80p062320,PRA79p052308}.
You et al. \cite{PRB81p014505} discussed a highly entangled ground state based on analyzing the coherent Hamiltonian without taking into account the dissipation.
Certain kinds of ground state entanglement in superconducting circuits  have been discussed in topological computation \cite{NaturePhys5p48} recently.
The possibility of ground state entanglement has also been discussed in the situation of a vanishing separation between two superconducting qubits \cite{PRA67p042319,PRB79p014516}. But in these references, neither the coherence nor the time evolution have been considered.
Experimentally, ground state entanglement with concurrence $C=0.33$ between two inductively coupled flux qubits coupled to a common heat bath has been observed \cite{PRL93p037003}. In this work, the always-on coupling (AOC) strength is close to the small transition frequency of qubits. This suggests that strong qubit-qubit interactions could be an interesting parameter regime for the creation and preservation of entanglement. This is further supported by related work on atoms subject to intense external driving fields, which can lead to a substantially modified incoherent dynamics of the system such as the population of excited bare states~\cite{PRL74p2451}.

Therefore, in this work,  we analyze a system of two interacting qubits coupled to a common bath. In contrast to previous results, we focus on the case of ultrastrong coupling between the two qubits, with coupling strength exceeding the respective qubit transition frequencies. 
We derive the corresponding master equation, interpret it in different collective state bases, and analyze the steady state properties and the temporal evolution without external driving fields, and with one or two external fields. 
The aim of the analysis is the generation of stationary entanglement between the two qubits.

We find that both the incoherent and the coherent dynamics is qualitatively modified by the strong always-on-coupling.  As a consequence, ground state entanglement can occur even in the absence of any driving field, with the entangled state populated by spontaneous emission only. By dynamically switching the strong qubit-qubit coupling on and off, we found that starting from an initially unentangled ground state of the two qubits, entanglement with high fidelity can be generated on time scales on the order of the inverse qubit transition frequency. With the help of  a single external driving field, both symmetric and antisymmetric entangled  qubit states can be prepared and maintained with  high fidelity. Finally, using two external fields, entangled states can be trapped based on so-called dark-states or prepared via adiabatic passage methods. Possible applications of our scheme include nonlocal quantum gates and quantum memories.


\section{\label{master}Master equation}
\subsection{Model}
Our system consists of two flux qubits coupled to each other through their mutual effective inductance $M$ and to a reservoir of harmonic oscillators modeled as LC oscillators, as shown in Fig.~\ref{fig:system}. Both qubits are operated as two-level quantum systems. A crucial difference to previous work~\cite{PRL96p067003,PRB79p024519} is that in the present work, ultra-strong coupling due to a large mutual inductance $M$ is considered. This coupling can be engineered~\cite{PRL94p090501,PRL96p047006}, and is dynamically controllable~\cite{Science316p723,PRL98p057004,PRB79p020507R,PRB70p140501R,PRB74p184504,PRB73p094506,APL83p2387}. In particular, it can be comparable to~\cite{PRL94p090501,PRL98p057004} or even much larger~\cite{PRL96p047006,PRB74p220503R,PRB72p020503R} than the qubits' transition frequencies. 

Each qubit is modeled as a loop interrupted by three Josephson junctions: two identical junctions characterized by the ratio of Josephson energy to charging energy $E_J^{(l)}/E_C^{(l)}$, and a third one with area smaller by a factor $\xi_l$ than the other two. Here, $l\in \{1,2\}$ labels the two qubits. Thus, the Josephson energies and capacitances in the $l$th qubit loop are given by
\begin{subequations}
\begin{align}
E_{J1}^{(l)}&=E_{J2}^{(l)}=E_{J}^{(l)},
\qquad E_{J3}^{(l)}=\xi_l E_{J}^{(l)} \,,\\
C_{J1}^{(l)}&=C_{J2}^{(l)}=C_{J}^{(l)}
, \qquad  C_{J3}^{(l)}=\xi_l C_{J}^{(l)}\,.
\end{align}
\end{subequations}
The gauge-invariant phase drops across the three junctions in the $l$th qubit are $\phi_{1}^{(l)},\phi_{2}^{(l)}$ and $\phi_{3}^{(l)}$.  
The qubits are modeled as two-level system, and in pseudo-spin operator language, the single-qubit Hamiltonian can be written as
\begin{equation}\label{eq:SHQ}
 H_Q^{(l)}=\frac{1}{2}\left[ t_l\sigma_x^{(l)}+\epsilon_l\sigma_z^{(l)}\right]\,.
\end{equation}
This model is valid for both three- or four- junction flux qubits~\cite{PRB73p094506,PRB60p15398}. The transition frequencies $\epsilon_l=2I_p^{(l)}(\Phi_e^{(l)}-\Phi_0/2)$ are tunable via individual static bias magnetic fluxes $\Phi_e^{(l)}$ characterized by the bias $f_b^{(l)}=\Phi_e^{(l)}/\Phi_0$, where  $\Phi_0=h/(2e)$ is the superconducting flux quantum. The persistent current through qubit $l$ is $I_p^{(l)}$, and the tunnel coupling between the two potential wells is denoted as $t_l$.
Additionally, each qubit is driven by a TDMF
\begin{align}
\Phi^{(l)}_e(t)=A_l\, \cos(\omega_c^{(l)} t)\,,
\end{align}
with amplitude $A_l$ and frequency $\omega_c^{(l)}$.
%
\begin{figure}[t]
 \centering
 \includegraphics[width=0.9\columnwidth]{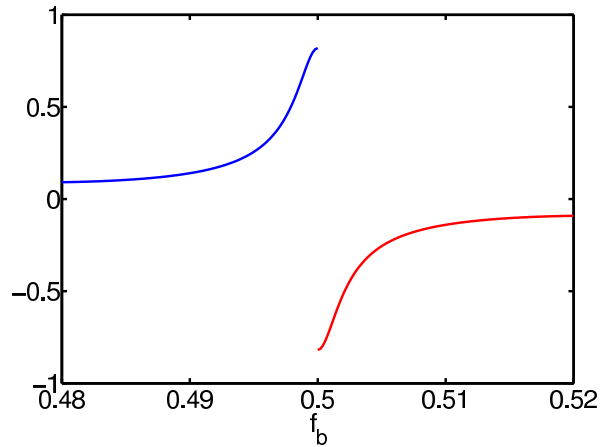}
 \caption{\label{fig:dipole}(Color online)  The transition dipole matrix element as a function of bias $f_b=\Phi_e/\Phi_0$ for $\xi=0.8, E_J/E_C=35$.}
\end{figure}
%
The coupling between the two qubits is modeled using a generic Ising-type coupling
\begin{align}
J\sigma_z\otimes \sigma_z\,.
\end{align}
We operate in the small vicinity of the optimum point $f_b^{(l)}=0.5$, and use a rotated coordinate system such that $\sigma_z$ and  $\sigma_x$ are effectively exchanged~\cite{PRB60p15398,arxiv10051559}. The coupled qubit pair driven by the TDMFs can then be described by the following Hamiltonian in a bare-state basis~\cite{PRL96p067003,PRB79p024519}
\begin{align}\label{eq:HQ}
    H_Q& =\frac{1}{2} \sum_{l=1}^2 {\hbar \omega_0^{(l)} \sigma_z^{(l)}} \nonumber\\
     & \quad -\hbar\sum_{l=1}\left (\sigma_+^{(l)}e^{-i\omega_ct}+H.c. \right ) \left ( k_l e^{-i\omega_c^{(l)} t}+c.c \right )\nonumber\\
     & \quad -\hbar\sum_{l\neq m=1}^2
     \left (\sigma_+^{(l)}\sigma_-^{(m)}+H.c. \right )
\left (\Omega_{lm}^{(1)}e^{i\omega_c^{(l)} t}+c.c \right )\nonumber\\
     & \quad -\hbar\sum_{l\neq m=1}^2
     \left (\sigma_+^{(l)}\sigma_+^{(m)}+H.c. \right )
\left (\Omega_{lm}^{(2)}e^{i\omega_c^{(l)} t}+c.c \right )\nonumber\\
     & \quad +\hbar \left (J\sigma_+^{(1)}\sigma_-^{(2)}+J\sigma_+^{(1)}\sigma_+^{(2)}+H.c. \right )\,.
\end{align}
The transition frequency $\omega_0^{(l)}$ of qubit $l$ is determined by
\begin{align}
\hbar\omega_0^{(l)}=\sqrt{t^2_l+\epsilon_l^2}\,.
\end{align}
The Pauli spin matrices $\sigma_{+,-,z}$ of the $l$th qubit with ground state $|g_l\rangle$ and excited state $|e_l\rangle$ are defined as 
\begin{subequations}
\begin{align}
\sigma_z^{(l)}&=|e_l\rangle\langle e_l|-|g_l\rangle \langle g_l|\,,\\
\sigma_+^{(l)}&=|e_l\rangle \langle g_l|\,,\\ \sigma_-^{(l)}&=|g_l\rangle \langle e_l|\,.
\end{align}
\end{subequations}
Introducing 
\begin{subequations}
\begin{align}
\phi_p^{(l)} &= (\phi_1^{(l)}+\phi_2^{(l)})/2 \,,\\
\phi_m^{(l)} &= (\phi_1^{(l)}-\phi_2^{(l)})/2 \,,
\end{align}
\end{subequations}
the coupling strengths can be written as~\cite{PRL96p067003}
\begin{subequations}
\begin{align}
k_l &= A_l \langle e_l|(I_p^{(l)}+i\Upsilon_l P_{P,l}) |g_l \rangle/(2\hbar) \,,\\
\Omega_{lm}^{(1)}&=A_l\beta_l \langle e_l, g_m|I_p^{(m)}\cos(2\phi_p^{(l)}+2\pi f_b^{(l)} ) |g_l, e_m\rangle /2\,,\\
\Omega_{lm}^{(2)} &=A_l\beta_l \langle e_l, e_m|I_p^{(m)}\cos(2\phi_p^{(l)}+2\pi f_b^{(l)} ) |g_l, g_m\rangle /2 \,,
\end{align}
\end{subequations}
and can be controlled by the applied TDMFs. Here, the parameters are given by~\cite{PRL96p067003}
\begin{subequations}
\begin{align}
\Upsilon_l &= \frac{2\pi \xi_l \omega_c^{(l)}}{(1+2\xi_l) \Phi_0} \,,\\
\beta_l &= M\left(\frac{2\pi}{\Phi_0}\right)^2 \frac{\xi_l  E_{J}^{(l)}}{2(1+2\xi_l)} \,,
\end{align}
\end{subequations}
the effective momenta are
\begin{equation}
 P_{P,l}=-i\hbar\frac{\partial}{\partial \phi_p^{(l)}}\,,
\end{equation}
 and the dipole moments are defined as  \cite{PRL93p087003}
\begin{align}
\vec{d}_l &= \frac{\Phi_0 I_p^{(l)}}{2\pi E_J^{(l)}} \,\\ \nonumber
          &= \langle e_l | \sin(2\phi_p^{(l)}+2\pi f_b^{(l)}) |g_l\rangle \,.
\end{align}
The always-on  coupling (AOC) energy arising from the mutual inductance induced qubit-qubit coupling is given by~\cite{PRL96p067003,PRL93p037003}
\begin{equation}\label{eq:AOC}
 J=MI_p^{(1)}I_p^{(2)} \,.
\end{equation}
This AOC strength is real in the vicinity of the optimum point $f_b^{(l)}=0.5$. The ultra-strong coupling regime is characterized by $J$ exceeding the transition frequencies $\omega_0^{(l)}$ of the qubits.

Interestingly, the transition dipole matrix elements can change their signs as the bias fluxes sweep across the optimum points, see Fig.~\ref{fig:dipole}. We define the parameter
\begin{align}
p=\frac{\vec{d}_1\bullet \vec{d}_2}{|\vec{d}_1||\vec{d}_2|}
\end{align}
characterizing their relative orientation with $p=1$ denoting  parallel and $p=-1$ anti-parallel dipole moments.

\subsection{Diagonalizing the Hamiltonian}
As we focus on the strong coupling regime, a perturbation treatment~\cite{PRL96p067003,PRB79p024519,ZFicekQuantInterfCoh2004} of the coupling $J$ cannot be applied. Therefore, we transfer the system to a suitable dressed state basis, thus taking into account the AOC to all orders. The diagonalization of the Hamiltonian in Eqs.~(\ref{eq:HQ}) without TDMF but including the AOC leads to the eigenenergies $E_j$ and corresponding eigenstates $|j\rangle$  ($j\in\{1,2,3,4\}$), which can be interpreted as a single four-level system with states and energies. Note that the states and energies are different from the usual collective Dicke states commonly introduced for two weakly interacting two-level systems~\cite{ZFicekQuantInterfCoh2004}. Instead, due to the strong coupling, the eigenstates and energies are  given by
\begin{subequations}
\label{eq:DBasis}
\begin{align}
|1\rangle &= a|e_1,e_2\rangle+b|g_1,g_2\rangle\,,& E_1 &=\hbar \bar w\,, \\
|2\rangle &= \beta|e_1,g_2\rangle+\alpha|g_1,e_2\rangle\,,& E_2 &=\hbar w\,, \\
|3\rangle &= \alpha|e_1,g_2\rangle-\beta|g_1,e_2\rangle\,,& E_3 &=-\hbar w\,, \\
|4\rangle &= -b|e_1,e_2\rangle+a|g_1,g_2\rangle\,,& E_4 &=-\hbar \bar w\,,
\end{align}
\end{subequations}
with
\begin{subequations}
\begin{align}
\bar w&=\sqrt{\omega_0^2+J^2}\,, &\qquad w&=\sqrt{\Delta^2+J^2}\,,\\
D&=\omega_0+\bar w\,, &\qquad m&=\Delta+w\,,\\
a&=\frac{D}{\sqrt{D^2+J^2}}\,, &\qquad b&=\frac{J}{\sqrt{D^2+J^2}}\,,\\
\alpha&=\frac{m}{\sqrt{m^2+J^2}}\,, &\qquad \beta&=\frac{J}{\sqrt{m^2+J^2}}\,,\\
\omega_0 &= \frac{\omega_0^{(1)}+\omega_0^{(2)}}{2}\,, & \qquad
\Delta &= \frac{\omega_0^{(2)}-\omega_0^{(1)}}{2}\,.
\end{align}
\end{subequations}
In the following investigation, we assume two identical qubits, i.e., $\Delta=0$. This reduces the above expressions to $w=|J|$, $\alpha=\frac{1}{\sqrt{2}}$, $\beta=\frac{J}{\sqrt{2}|J|}$, and $a^2+b^2=1$. This assumption is reasonable as long as $\Delta\ll |J|$.
In terms of the new dressed-state basis, the coherent Hamiltonian of our system including the TDMFs becomes $H=H_0+H_I$ with 
\begin{subequations}
\label{eq:V}
 \begin{align}
  H_0=&\sum_j E_j R_{jj} \,,\\
  H_I=&\hbar \Omega_1 \left\{ (a\alpha+b\beta)(R_{12}+R_{34}) \right. \nonumber\\
&\quad  \left . +(a\beta+b\alpha)p e^{-i\phi_{21}} (R_{12}-R_{34})  \right\} e^{-i\omega_Lt} \nonumber\\ 
&+\hbar\Omega_2 \left\{ (a\beta-b\alpha)(R_{24}-R_{13}) \right. \nonumber \\
&\quad \left. +(a\alpha-b\beta)p e^{-i\phi_{21}} (R_{24}+R_{13}) \right\}e^{-i\omega_Lt}\nonumber\\ 
&+\hbar\Omega_3(a^2-b^2)\left(1+e^{-i\phi_{21}}\right)e^{-i\omega_Lt}R_{14}+ \textrm{H.c.}\,,
\end{align}
\end{subequations}
where the Pauli operators are defined as 
\begin{align}
R_{ij}=|i\rangle \langle j|
\end{align}
for $i,j\in\{1,2,3,4\}$. The interaction Hamiltonian $H_I$ arises from the coherent driving via the TDMFs. For simplicity, we have assumed that the field frequencies and Rabi frequencies satisfy $\omega_c^{(1)}=\omega_c^{(2)}=\omega_L$, $\Omega_1=\Omega_2=|k_1|=|k_2|$ and $\Omega_3=|\Omega_{12}^{(2)}|=|\Omega_{21}^{(2)}|$. The relative phase of the driving fields between the two qubits is 
\begin{align}
\phi_{21}=k_L r_{21}=2\pi (r_{21}/\lambda_0)(\omega_L/\omega_0)\,.
\end{align}

\subsection{Coupling to the bath}
We model the interaction of our qubit system with the environment via a reservoir of harmonic oscillators. The system-bath coupling can be described using the Jaynes-Cummings Hamiltonian~\cite{PRB79p024519,ZFicekQuantInterfCoh2004}
\begin{align}\label{eq:HB}
H_B& = \sum_k \hbar \omega_k a^{\dag}_k a_k -\hbar\sum_{l=1}^2\sigma_x^{(l)}
\sum_k \eta_k\left(e^{ikr_l} a_k+ \textrm{H.c.} \right)\nonumber\\
& \quad +\hbar\sum_k \sum_{l\neq m=1}^{2} \chi_k\left (e^{ikr_l} a_k+ H.c.\right ) \left (\sigma_+^{(l)}\sigma_+^{(m)}
+ \textrm{H.c.} \right )\,,
\end{align}
where $r_l$ is the position of $l$th qubit.
The bath oscillators have frequencies $\omega_k$, and $\langle a_k^\dag a_k \rangle=n_k$ is the average photon number of the $k$th field  mode. The thermal average number of photons for an oscillator of temperature $T$ is given by
\begin{align}
N_{th}(\omega)=[\exp(\hbar \omega/k_B T)-1]^{-1}\,,
\end{align}
and is assumed negligible at the frequencies relevant to our system. $\eta_r^{(l)}$ and $\chi_r^{(lm)}$ are determined by the vacuum field and are proportional to~\cite{PRL93p037003} $M_B^{(l)} I_p^{(l)}\sqrt{\hbar \omega_r/2L_B}$, where $M_B^{(l)}$ is the qubit-bath mutual inductance in the corresponding qubit. For simplicity, we assume $M_B^{(1)}=M_B^{(2)}=M_B$. In superconducting circuits, the qubits can be driven or biased by microwave circuits \cite{PRL96p127006,PRL93p037003,PRL94p090501,PRL96p047006} or $1$D transmission lines \cite{Science327p840,PRB78p180502R,PRA79p013819}. Thus the bath can be modeled as one-dimensional open space \cite{Science327p840}, and an Ohmic environment with a large cutoff $\omega_{cut}$ is a realistic assumption for this bath \cite{PRA67p042319,PRB65p144516}.

We assume that the coupling to the bath is weak, i.e., we operate in a Hamiltonian-dominated regime. Both qubits are operated near their optimal point, so that they are decoupled from low-frequency noise to a good degree, which otherwise would give rise to the pure dephasing. Then, applying the Born-Markov approximation to eliminate the bath \cite{PRB81p024520,PRB79p024519,PRA67p042319}, we find that the bath introduces both incoherent and coherent contributions to the dynamics of system via their dipole-dipole interaction (DDI), which we discuss next.

\subsubsection{Bath-induced incoherent contributions}
The incoherent part, i.e., the bath-induced dissipation, can be described by the Liouville operators $\mathscr{L}_{SP}\,\rho$ and $\mathscr{L}_{TP}\,\rho$ operating on the system density matrix $\rho$. The fist part $\mathscr{L}_{SP}\rho$ results from single-photon process proportional to $\eta_k$ in Eq.~(\ref{eq:HB}). It is composed of two contributions as
\begin{equation}\label{eq:Lrho}
 \mathscr{L}_{SP}\,\rho =\mathscr{L}_1\rho+ \mathscr{L}_2\,\rho\,,
\end{equation}
where 
\begin{subequations}
 \begin{align}
 \mathscr{L}_1{\rho}=&\mathscr{L}\{\gamma_{|1\rangle\rightarrow |2\rangle},R_{21}\}\rho+\mathscr{L}\{\gamma_{|2\rangle\rightarrow |4\rangle},R_{42}\}\rho \,,\nonumber \\ 
&+\mathscr{L}\{\gamma^{(cross)}_{|2\rangle},R_{12},R_{24}\}\rho\,,\\
 \mathscr{L}_2{\rho}=&\mathscr{L}\{\gamma_{|1\rangle\rightarrow |3\rangle},R_{31}\}\rho+\mathscr{L}\{\gamma_{|3\rangle\rightarrow |4\rangle},R_{43}\}\rho\nonumber \\
&+\mathscr{L}\{\gamma^{(cross)}_{|3\rangle},R_{13},R_{34}\}\rho\,,\\
 \mathscr{L}\{\kappa,A\}\rho=&\frac{\kappa}{2} \left(2 A \rho A^\dag-A^\dag A\rho- \rho A^\dag A\right)\,\\
 \mathscr{L}\{\kappa,A_1,A_2\}\rho =&\kappa (A_2^\dag \rho A_1+A_1^\dag \rho A_2)\,.
 \end{align}
\end{subequations}
The rates are given in Appendix~\ref{app-decay}. The decay rate of an isolate qubit is given by
\begin{equation}\label{eq:decayrate}
 \gamma_0(\omega)=\sin^2(\theta)(M_DI_p)^2\omega/(2\hbar Z) \,.
\end{equation}
This rate describes the damping of a qubit coupled to an $1$D open transmission line with impedance $Z=\sqrt{l_r /c_r}$ by a mutual inductance $M_B$~\cite{Science327p840}. Here, $l_r$ and $c_r$ are the inductance and capacitance per unit length, respectively.  In Eq.~(\ref{eq:decayrate}), $\sin(\theta)=t/\omega$ with tunneling $t$ and the transition frequency $\omega$. In our case, $\sin(\theta)\approx 1$. Note that an isolated qubit decays at a rate $2\gamma_0$ in our case. These results are in agreement with Ref.~\onlinecite{Science327p840}. The dipole-dipole crossing decay rate $\gamma_{12}$ evaluates to
\begin{equation}\label{eq:crossdamping}
 \gamma_{12}(\omega)=\gamma_0(\omega)\cos(\kappaup r_{21})\,,
\end{equation}
where $\kappaup=\omega/v$ is the wave number corresponding to the transition frequency and the wave phase velocity $v=1/\sqrt{l_r c_r}$. The effective separation between the two qubits is $r_{21}$. This formula was already presented in our previous work~\cite{PRB82p184532}, and a similar superradiant decay rate was  recently found in a donor-acceptor system mediated by plasmonic nano waveguides \cite{NanoLett10p3129,PRL106p020501}. 

The second part of the bath-induced decoherence $\mathscr{L}_{TP}\rho$ is due to two-photon processes and can be written as
\begin{equation}
\mathscr{L}_{TP}\rho = \mathscr{L}\{\gamma_{TP},R_{14}\}\rho\,,
\end{equation}
with the relaxation rate
\begin{equation}
 \gamma_{TP} =4(a^2-b^2)^2 (\tilde\gamma_0+\tilde \gamma_{12})\,,
\end{equation}
where the two-photon decay rate $\tilde{\gamma}_0$ is defined in Appendix~\ref{app-decay}. The rate $\tilde{\gamma}_{12}$ is given by
 \begin{equation}
  \tilde{\gamma}_{12}=\tilde{\gamma}_0 \cos(\kappaup r_{21})\,.
 \end{equation}
All relaxation rates are derived in Born-Markov approximation neglecting small Lamb shifts~\cite{ZFicekQuantInterfCoh2004,PRA2p883}, and are summarized in Appendix~\ref{app-decay}. 

\subsubsection{Bath-induced coherent contributions}
The coherent part $H_{CDDI}$ consists of dipole-dipole energy shifts
\begin{align}
H_{DDS}=\hbar \sum_j\omega_j^{(DDS)}|j\rangle \langle j|
\end{align}
in the transition frequencies and of the bath-induced excitations $H_{BIE}$. These two contributions are discussed in detail in Appendix~\ref{app-dd}.

\subsection{Master equation}

After including the incoherent and coherent contributions,
the dynamics of our system can be described by a master equation as
\begin{align}
\label{eq:MEq}
 \frac{\partial \rho}{\partial t}=&-i\sum_j \omega_j[R_{jj},\rho]-\frac{i}{\hbar}\left[H_{I}+H_{BIE},\rho\right]\nonumber \\
&+\mathscr{L}_{SP}\rho+\mathscr{L}_{TP}\rho+ \mathscr{L}_\phi \rho\,.
\end{align}
Here, we redefined the transition frequency to also include the dipole-dipole shifts as
\begin{equation}\label{eq:DDS}
 \hbar\omega_j=E_j+\hbar\omega^{(DDS)}_j\,.
\end{equation}
The final term $\mathscr{L}_\phi$ phenomenologically introduces pure dephasing with rate $\Gamma_\phi$ for the two qubits to the master equation Eq.~\ref{eq:MEq}, and is given by
\begin{equation}
 \mathscr{L}_\phi \rho=\sum_j \mathscr{L}\left\{\Gamma_\phi/2,|e_j\rangle \langle e_j|-|g_j\rangle \langle g_j|\right\}\rho\,.
\end{equation}
Note that this treatment of pure dephasing is only valid for operation near the optimum point, where the qubit in leading order decouples from fluctuations in the external fields, and thus becomes insensitive to dephasing induced, e.g., by $1/f$ noise.

\subsection{Model parameters}
As already mentioned, we assume identical qubits and equivalent TDMF parameters for the two qubits. Furthermore, for our calculations, we assume a large two-photon decay rate $\tilde{\gamma}_0=0.02\gamma_0$ and $r_{21}$ between $\lambda_0/1000$ and $\lambda_0/50$ corresponding to an effective separation of around $10-200$ {\micro\meter} between the two qubits \cite{PRL94p090501,PRB67p220506R}.
Note that the separation $r_{21}/\lambda$ scaled by the wavelength even can be dynamically changed in superconducting qubits because the transition frequency $\omega$ is tunable via the bias $f_b$. It also can be influenced  by auxiliary qubits \cite{PRL98p057004,PRB79p020507R}.
The qubit transition frequencies are taken as $\omega_0=1000\gamma_0$,  and the bath cutoff as  $\omega_{cut}=200\omega_0$. We adjust the bias fluxes to operate the qubits near their optimal points $f_b^{(k)}=0.5$ and assume identical, but uncorrelated pure dephasing for the two qubits. 

\subsection{Experimental realization}
We now discuss aspects of the experimental realization of our framework based on existing experiments and proposals. As our analysis focuses on the case of strong coupling between the qubits, it is important to note that such a strong AOC between two flux qubits comparable or even larger than their transition frequencies has been realized~\cite{PRL94p090501,PRL98p057004,PRL96p047006,PRB74p220503R,PRB72p020503R}. This coupling can also be dynamically controlled in experiments~\cite{Science316p723,PRL98p057004,PRB79p020507R,PRB70p140501R,PRB74p184504,PRB73p094506,APL83p2387}. These advances open the door to the study of an entirely new parameter regime and motivate our work.

Regarding the qubits themselves, in practice, two fabricated qubits will always be slightly different,  e.g., in the transition frequency at the optimal point. But this effect of nonidentical qubits is to a good degree negligible since the coupling of interest in our work is at least comparable to the transition frequency, and thus strongly dominates over possible detunings of the two qubits.

Next, we discuss the bath, and focus in particular on its 1D nature, and the fact that we treat the qubits as coupled to a common bath. In our system,  the superconducting qubits are driven by microwave fields led by microwave lines~\cite{PRL97p167001,PRL98p257003,PRB81p024520,PRB76p214503} or transmission lines~\cite{Science327p840,PRL101p080502,PRL103p083601}. In both cases, the spontaneous emission is the dominant source of energy relaxation~\cite{PRL97p167001,PRB81p024520,PRB76p214503,Science327p840,PRL101p080502}, which results mainly from high-frequency noise.
In a recent experiment \cite{Science327p840}, a flux qubit is coupled to a 1D transmission line. The qubit is well isolated from the other degrees of freedom in the surrounding solid-state environment and behaves as a nearly isolated artificial atom in open space, coupled only to the  electromagnetic fields in the 1D space with long correlation length. In this work, the authors presented an analytical expression for the relaxation rate which is in good agreement with the experimental observation. 
Storcz and Wilhelm \cite{PRB67p042319} provide two flux qubit coupled by flux transformers \cite{Science285p1036} as an example for a situation described by a common bath. 
Note that both microwave lines and transmission lines induce an Ohmic environment and can be modeled as a global LC oscillator \cite{PRB81p024520,PRB76p214503,Science327p840,PRL103p083601} as in our analysis. Thus the above works indicate possibilities to implement  our framework.

We have furthermore assumed a bath of negligible temperature at the qubit transition frequencies. 
A bath of finite temperature would gives rise to an incoherent pumping to the excited state of system. Thus, the system would evolve into a mixed state in thermal equilibrium even if no driving field is applied. This can lead to a reduction of the entanglement found in some of the cases discussed below. Thermal pumping can best be avoided by using qubits with large transition frequency embedded in a low-temperature refrigerator.

Next, we turn to the effects of the dipole-dipole interaction, which arise from the coupling to a common bath. Due to this coupling, two qubits decaying to a common bath may exhibit  collective effect such sub- and superradiant states.
Recent experiments~\cite{Nature449p443,PRL103p083601} have observed such superradiant and subradiant states of two distant superconducting qubits coupled via cavity modes. The dipole-dipole interaction of the qubits in these two works is a result of exchange of virtual  photons via the cavity. This interaction is also the origin of the collective effects. This cavity can be replaced by flux transformers \cite{Science285p1036} or a common readout or driving circuit \cite{PRL93p037003,QIP8p133}.
Two adjacent flux qubits coupled by magnetic dipole-dipole interactions are also discussed in a recent review by Clarke and Wilhelm~\cite{Nature453p1031}. Finally, the DDI between two flux qubits can also be induced by flux transformers \cite{Science285p1036} or a quantum bus \cite{PRL103p083601,Nature449p443}. 

Finally, we discuss decoherence. The decoherence consists of two contributions. The first contribution arises from the relaxation process, and the second is pure dephasing arising typically  from low-frequency noise such as $1/f$-noise.  Our system is operated near the optimum point $f_b=0.5$ of the flux qubits, in order to reduce the effect of the pure dephasing. Then,  the qubits  to first order are decoupled from the dephasing fluctuations~\cite{Nature453p1031,PRL97p167001,PRB76p214503}. This allows us to phenomenologically treat the pure dephasing via a Markovian master equation.

\section{\label{drive}Results}
\subsection{\label{dme}Discussion of master equation}
In this part we discuss some of the properties of the master equation~(\ref{eq:MEq}), and compare it to corresponding master equations in atomic or ionic systems. We find that the implementation in strongly coupled qubits offers a number of distinct features. This mainly arises from the fact that in our model, the superconducting qubits are not only coupled via the bath, but they can also be directly coupled via their mutual inductance. This coupling includes a two-photon channel, which is not present in typical atomic/ionic systems. Furthermore, we consider the ultra-strong coupling regime, which appears unfeasible in atomic systems
\begin{figure}[t]
\centering
\includegraphics[width=0.9\columnwidth]{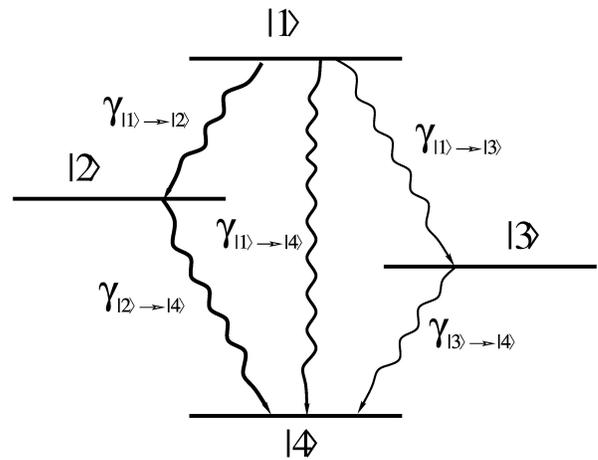}
\caption{\label{fig:DL}(Color online) The relevant relaxation channels in the collective dressed state basis. The transition $|1\rangle \to |4\rangle$ is a two-photon channel.}
\end{figure}

\begin{figure}[t]
\centering
\includegraphics[width=0.9\columnwidth]{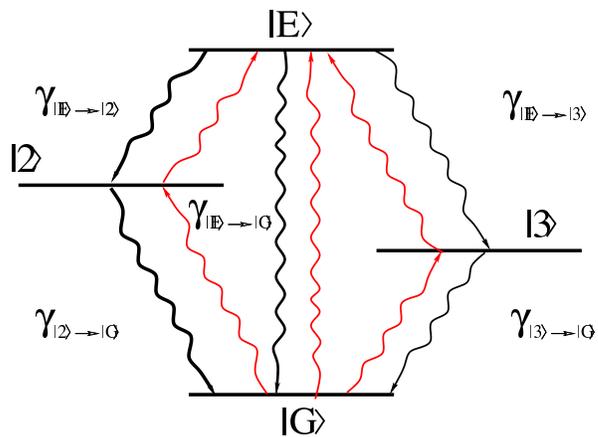}
\caption{\label{fig:BL}(Color online) The system in the collective Dicke state basis. The black downwards arrows denote dissipation, whereas the red upwards arrows indicate pumping which is introduced in the collective state basis due to relaxation in the dressed-state basis shown in Fig.~\ref{fig:DL}.}
\end{figure}

We first note that the dissipative dynamics is strongly modified by the strong AOC via the mutual inductances. In atomic systems, the relaxation is determined by the collective dynamics depending on the separation of the two particles and their mutual dipole orientation~\cite{ZFicekQuantInterfCoh2004}. In our system, the effect of dissipation can most easily be described in the dressed state basis shown in Fig.~\ref{fig:DL}, and defined in Eq.~(\ref{eq:DBasis}). According to the master equation, as expected, the upper states decay to the lower states, such that all population resides in the ground state $|4\rangle$ in steady state in the absence of a driving field. However due to the strong AOC, this ground state can have a rather complex structure. To establish this result, we transfer our system into the collective Dicke state basis with upper state $|E\rangle=|e_1,e_2\rangle$ and lower state $|G\rangle=|g_1,g_2\rangle$ as shown in Fig.~\ref{fig:BL}. This basis is frequently used to analyze weakly coupled two-qubit systems.  In this representation, the coupling to the environment not only leads to relaxation from states with higher energy to states with lower energy,  but also effectively to pumping from lower to higher states, and even to some coherent interaction (see Appendix~\ref{app-shift}). Interestingly, both damping and pumping emerge in every transition channel, even though the pumping is always smaller than the damping $(|b|<|a|)$. Thus we can conclude that already the dissipative dynamics can lead to the formation of states of interest, and we will show below that this also includes highly entangled states.

\begin{figure}[t]
 \centering
 \includegraphics[width=0.9\columnwidth]{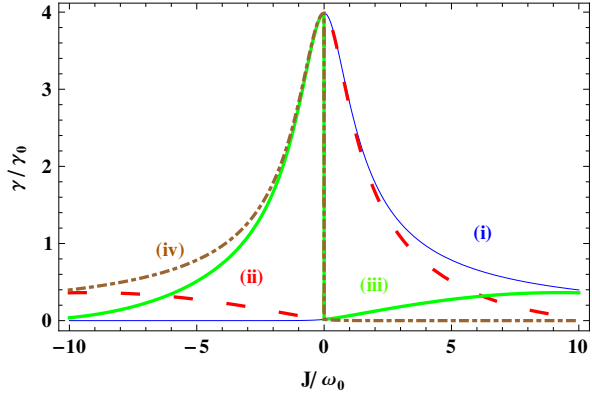}
 \caption{\label{fig:relJ}(Color online) The collective decay rates  $\gamma_{1\rightarrow 2}$ [(i) thin blue line], $\gamma_{2\rightarrow 4}$ [(ii) red line], $\gamma_{1\rightarrow 3}$ [(iii) green line] and $\gamma_{3\rightarrow 4}$ [(iv) brown line] as a function of the AOC coupling $J$. The distance is chosen as $r_{12}=\lambda_0 /50$. In (a), the two transition dipole moments are parallel ($p=1$). The result for anti-parallel dipole moments ($p=-1$) is obtained by mirroring the figure at the $J=0$ axis.}
\end{figure}

\begin{figure}[t]
 \centering
 \includegraphics[width=0.9\columnwidth]{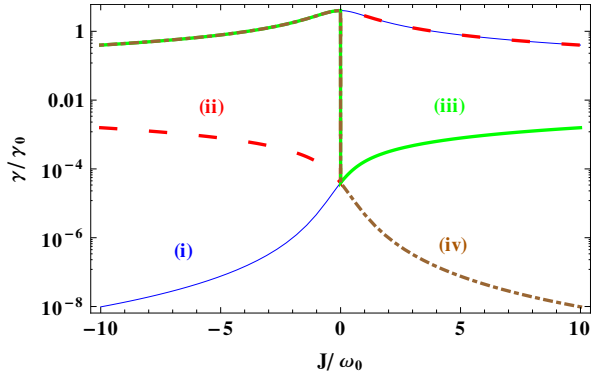}
 \caption{\label{fig:relJsmallR}(Color online) The collective decay rates  $\gamma_{1\rightarrow 2}$ [(i) thin blue line], $\gamma_{2\rightarrow 4}$ [(ii) red line], $\gamma_{1\rightarrow 3}$ [(iii) green line] and $\gamma_{3\rightarrow 4}$ [(iv) brown line] as a function of the coupling $J$ for $r=\lambda_0/1000$ and $p=1$.}
\end{figure}

We now turn to a discussion of the collective state decay rates indicated in Fig.~\ref{fig:DL}. In the weak-coupling regime, these cooperative relaxation rates are essentially determined by the effective separation $r_{21}$ between the two qubits and their transition frequencies. But in our model system, the relaxation rates in addition crucially depend on the strong AOC between the qubits. As an example, we show the dependence of the collective decay rates on the AOC strength $J$ in Fig.~\ref{fig:relJ}. For small $J\ll \omega_0$, the relaxation rates in each channel are equal. In contrast, for large AOC $J\geq \omega_0$, the four relaxation rates become substantially different. As an example, we note that the decay from the excited state $|1\rangle$ to the intermediate state $|3\rangle$ is much faster than that of $|3\rangle$ to $|4\rangle$ if the AOC is strong and if the coupling constant $J$  has the same sign as $p$. This indicates a possibility to trap the system in the intermediate state $|3\rangle$, which can either be a symmetric or an antisymmetric state depending on the sign of the AOC constant $J$, see Eq.~(\ref{eq:DBasis}). 
Interestingly, the rate $\gamma_{|1\rangle \rightarrow |3\rangle}$ can be comparable to $\gamma_{|1\rangle \rightarrow |2\rangle}$. However, within a practical AOC strength $J$, this is unfeasible  for the small separation $r_{21}=\lambda_0/1000$ shown in Fig.~\ref{fig:relJsmallR}.

%

We note that related effects due to strong coherent driving have already in detail been discussed in a number of previous works in atomic systems subject to strong external driving fields~\cite{PRL74p2451,PRL91p123601,PRL89p163601,PRL83p1307}, and recently also in superconducting circuits strongly driven by a microwave field~\cite{PRB81p024520,PRL105p257003}.

\begin{figure}[t]
 \centering
 \includegraphics[width=0.9\columnwidth]{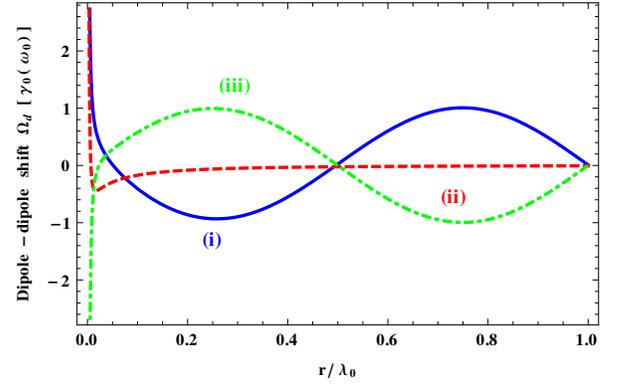}\\
 \caption{\label{fig:DDS}(Color online) The dipole-dipole induced shifts $\Omega_{d}^+$ [(i) blue line], $\Omega_{d}^-$ [(ii) dashed red line] and $\Omega_{d}$ [(iii) dash-dotted green line] as a function of the qubit separation $r_{21}/\lambda_0$ at a fixed frequency $\omega_0$.}
\end{figure}

It is also interesting to discuss the role of pure dephasing in this context. In the dressed state basis, the decoherence from pure dephasing becomes
\begin{align}
 \mathscr{L}_\phi \rho =& \sum_j \mathscr{L}_j \{\Gamma_\phi/2,|e_j\rangle \langle e_j|-|g_j\rangle \langle g_j|\}\rho \nonumber\\
=& \mathscr{L}\{(a^2-b^2)^2 \Gamma_\phi,|1\rangle \langle 1|-|4\rangle \langle 4| \}\rho \nonumber\\
&+\mathscr{L}\{4a^2b^2\Gamma_\phi,|1\rangle \langle 4|+|4\rangle \langle 1|\}\rho \nonumber\\
&+\mathscr{L}\{\Gamma_\phi,|2\rangle \langle 3|+|3\rangle \langle 2|\}\rho  \nonumber\\
&-\mathscr{L}\{-2ab(a^2-b^2)\Gamma_\phi,|1\rangle \langle 1|-|4\rangle \langle 4|,\nonumber \\
& \qquad |1\rangle \langle 4|+|4\rangle \langle 1|\}\rho\,.
\end{align}
We find that independent of the AOC strength, the pure dephasing induces incoherent energy exchange between two intermediate states $|2\rangle$ and  $|3\rangle$. However, it gives rise to incoherent energy exchange between $|1\rangle$ and  $|4\rangle$ only if the AOC is large. For small AOC the pure dephasing rates of $|1\rangle$ and  $|4\rangle$ are equal to those of an isolated qubit. This part becomes small if the AOC is large.

Next, we analyze the effect of the dimensionality of the environment.
In the atomic model of coupled qubits, typically a 3D reservoir is considered~\cite{PRA2p883}. Then, the dipole-dipole shifts (DDS) $\Omega_d=-\left(\Omega_d^{+}+\Omega_d^{-}\right)$ for small distances $r_{21}$ decay as $r_{21}^{-3}$ and for larger distances vanish with oscillations around zero and an overall decay proportional to $r_{21}^{-1}$. 
In our model, the reservoir effectively forms a one-dimensional open space \cite{Science327p840}. In 1D, the DDS $\Omega_d$ is much smaller for small separation $r_{21}<\lambda_0/10$ than in 3D, see Fig.~\ref{fig:DDS}. The DDS $\omega_2^{DDS}-\omega_3^{DDS}$ between the two intermediate levels $|2\rangle$ and  $|3\rangle$ quickly decays proportional to  $r_{21}^{-1}$ as the separation increases to $r_{21}>\lambda_0/50$. This shift is about $100\gamma_0$  for $r_{21}=\lambda_0/1000$ but decreases to less than $\gamma_0$ if $r_{21}\leq\lambda_0/50$. Furthermore, $\Omega_d$ reduces to a sinusoidal oscillation if $r_{21}>\lambda_0/10$, as demonstrated for two qubits with coupling mediated by one-dimensional plasmonic waveguides \cite{PRL106p020501}.  Thus, the DDS effects on the evolution of system is small except for trivial coherent shifts in the transition frequencies. We note that our results is different from the prediction in~\cite{PRL106p020501} in the case of a distance much smaller than the wavelength. Our model calculates the principal value integration for the DDS, whereas in~\cite{PRL106p020501}, an approximation involving Green's function was used. In our case, the DDS in 1D space exponentially decays. This decay is similar to the 3D case in which the loss from diffraction is also negligible if the emission from the donor directly points to the acceptor.

\begin{figure}[t]
\centering
\includegraphics[width=0.9\columnwidth]{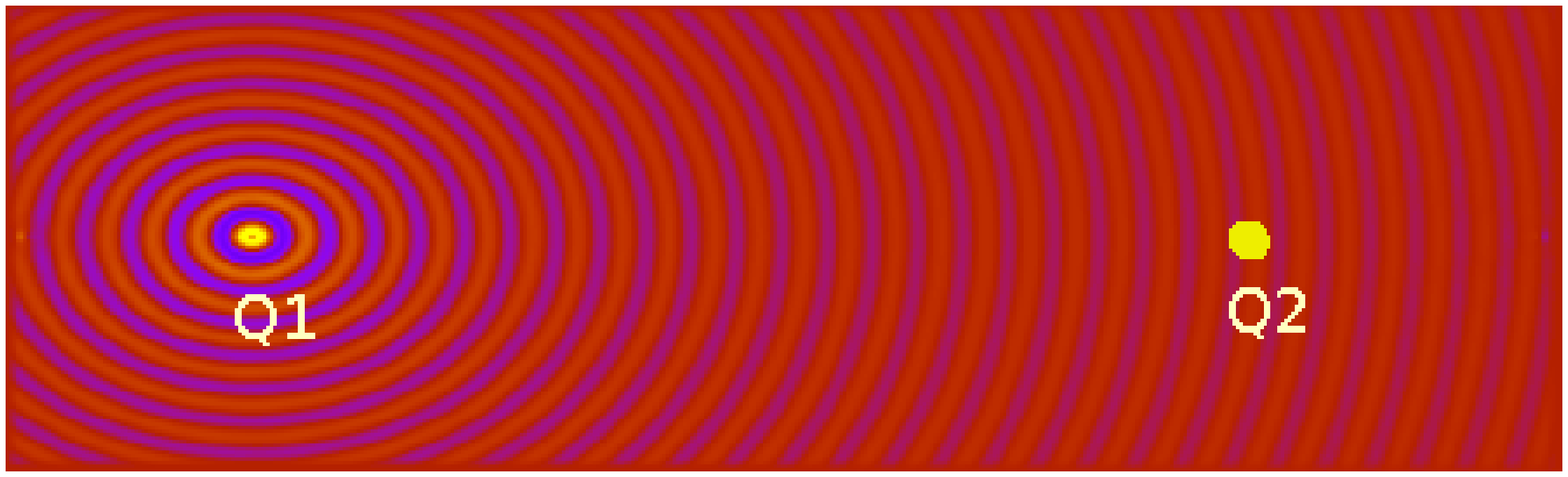}\\
\includegraphics[width=0.9\columnwidth]{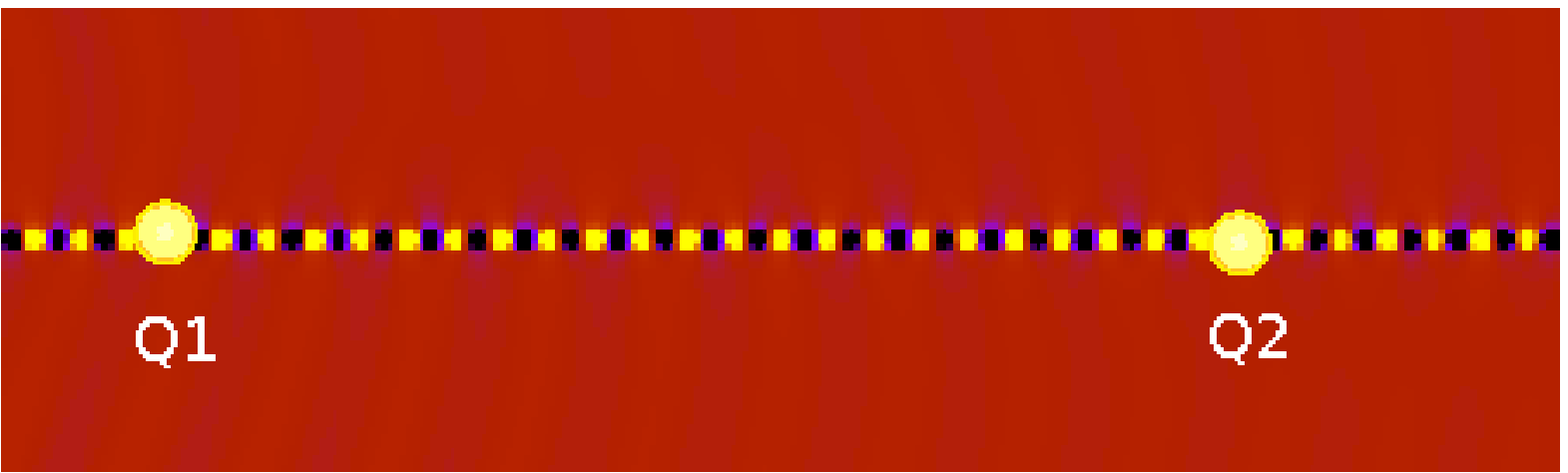}
\caption{\label{fig:rel}(Color online) (a) Pictorial representation of a photon emission from qubit Q1 in an open 2D space. (b) The corresponding situation in $1$D space. The results are obtained by numerically solving Maxwell's equations for an oscillating source with the finite-difference time-domain method (FDTD) \cite{FDTD}.}
\end{figure}

%
%
%
%
Also the collective relaxation rates of the qubit in 1D are essentially different from the higher dimensional case. In the 3D case, the decay cross rate $\gamma_{12}$ tends to $\gamma_0$ for small separations $r_{21}\ll \lambda_0$. For larger distances, it oscillates and overall decays rapidly  as the separation $r_{21}$ increases~\cite{PRA2p883}. In contrast, in an 1D bath, this cross damping rate oscillates as a  $\cos$ function of the separation $r_{21}$, but does not have an overall decay with increasing distance, see Eq.~(\ref{eq:crossdamping}). This can easily be understood by noting that in 1D, the coupling Hamiltonian essentially does not carry distance information, and photons emitted from one atom are effectively guided to the other atom via the 1D space. In contrast, in 3D, photons have a high probability to miss the other atom, as shown in a rather pictorial way in Fig.~\ref{fig:rel}. This feature is comparable to atoms embedded in a suitable cavity, which also can render the environment predominantly one-dimensional.

\subsection{Ground state entanglement}
Creating and protecting high entanglement in a controllable way is of interest, as the decoherence arising from the coupling of a system to its environment tends to rapidly destroy entanglement. This is particularly difficult in solid-state superconducting circuits.

In the following, we first demonstrate the possibility to create ground state entanglement based on the modification of dissipation due to the strong coupling between the qubits. In particular, no external coherent driving is required, and we neglect the corresponding Hamiltonian $H_I$. Thus we will show that in our system, the dissipation can be a useful resource for quantum entanglement without any coherent driving field~\cite{PRL85p1762}. Note that the usefulness of dissipation has recently also been demonstrated in a carefully engineered local environment~\cite{NaturePhys5p633}.  The possibility of ground state entanglement in superconducting qubits has been discussed for vanishing distances between two qubits~\cite{PRA67p042319,PRB79p014516}. However, in this study, neither the operation time  nor the collective coherent driving induced by by dissipation was considered.

\begin{figure}[t]
 \centering
 \includegraphics[width=0.9\columnwidth]{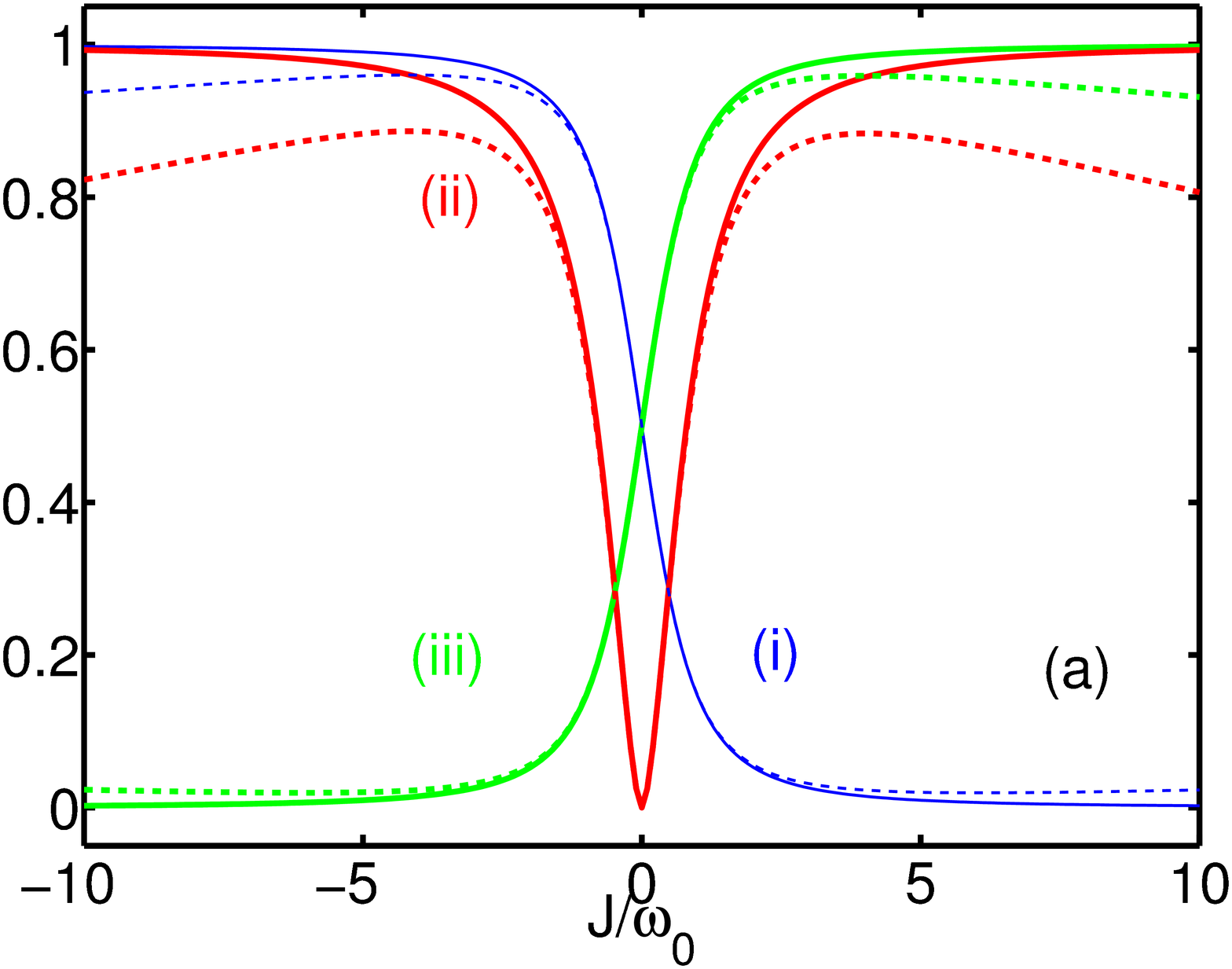}\\
\includegraphics[width=0.9\columnwidth]{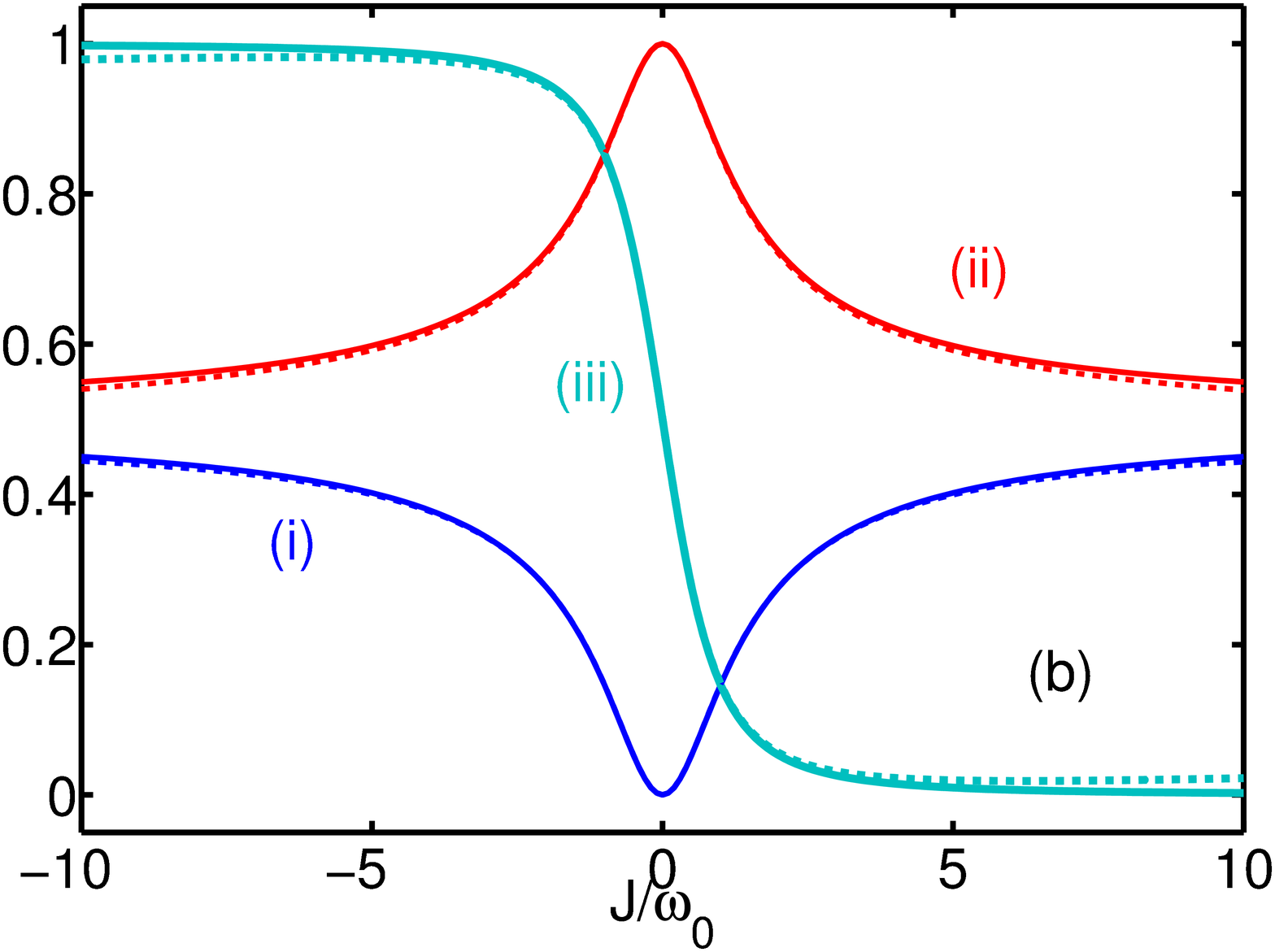}
 \caption{\label{fig:GSE}(Color online) (a) Fidelities of the Bell states $|\Phi_+\rangle$ [thin blue line (i)], $|\Phi_-\rangle$ [green lines (iii)], and entanglement [red lines (ii)] as a function of the AOC $J$. 
(b) Fidelities of the product states $|e_1,e_2\rangle$ [blue lines (i)], $|g_1,g_2\rangle$ [green lines (ii)], and coherence [thick cyan lines (iii)] as a function of the AOC $J$. 
In both subpanels, the parameters are $p=1$ and $r=\lambda_0/1000$. Note that the coherence was shifted upwards by $0.5$. The solid lines indicate the case without pure dephasing, while the dashed lines include a pure dephasing of $\Gamma_\phi=0.01\gamma_0$.}
\end{figure}

As discussed in Sec.~\ref{dme}, in the dressed state basis all upper states of our system decay to the ground state, see Fig.~\ref{fig:DL}. We have verified this by numerically solving the master equation (\ref{eq:MEq}) with $H_I=0$, as shown in Fig.~\ref{fig:GSE}(a).
To assess the ground state entanglement, we calculate the entanglement of formation of the two qubits in the magic basis as~\cite{PRL80p2245}
\begin{equation}
  E(\rho)=\varXi (C(\rho))\,,
\end{equation}
where
\begin{subequations}
 \begin{align}
  \varXi (C)&=h\left (\frac{1+\sqrt{1-C^2}}{2}\right )\,\\
  h(x)&=-x\,\log_2x-(1-x)\log_2(1-x)\,,
 \end{align}
\end{subequations}
 with concurrence $C(\rho)=max\{0,\lambda_1-\lambda_2-\lambda_3-\lambda_4\}=max \{0,2\lambda_{max}-Tr[R]\}$. The $\lambda_i$ are the eigenvalues, in decreasing order, of the Hermitian matrix $R$ given by
\begin{subequations}
\begin{align}
R&=\sqrt{\sqrt{\rho}\tilde \rho \sqrt{\rho}}\,,\\
\tilde \rho&=(\sigma_y \otimes\sigma_y )\rho^*(\sigma_y \otimes\sigma_y )\,.
\end{align}
\end{subequations}
For small coupling strengths $J\ll \omega_0$, our system reduces to the usual configuration studied, e.g., in~\cite{PRL96p067003,PRB79p024519}. In essence, the ground state tends to a product state, and neither entanglement nor coherence occurs, see Fig.~\ref{fig:GSE}.

Next, we focus on the case of large AOC, both without pure dephasing and with pure dephasing of $\Gamma_\phi=0.01\gamma_0$.
Surprisingly, for the AOC $|J|$ larger than the resonance frequency of the qubits, the ground state tends to a Bell state 
\begin{align}
|\Phi_\pm\rangle= \frac{1}{\sqrt{2}}(|g_1,g_2\rangle\pm|e_1,e_2\rangle)
\end{align}
with large fidelity, such that the entanglement becomes high. The stronger the AOC is, the higher is the fidelity of the Bell state, and subsequently the larger the entanglement.
For a large coupling $|J|\geq 5\omega_0$, the fidelity of the Bell states exceeds $97\%$ in the steady state if no pure dephasing occurs, as shown by the solid lines in Fig.~\ref{fig:GSE}(a). The ground state is close to $|\Phi_-\rangle$ for a positive $J$, while it becomes $|\Phi_+\rangle$ if $J<0$.
By dynamically tuning the sign of the coupling~\cite{Science316p723,PRL98p057004,PRB79p020507R,PRB70p140501R,PRB74p184504,PRB73p094506,APL83p2387}, the ground state can thus be switched between the two Bell states $|\Phi_\pm\rangle$. Alternatively, the bias flux can be inverted around the optimal point because this also inverts the sign of the AOC, see Fig.~\ref{fig:dipole}. As expected, the population in the product states $|g_1,g_2\rangle$ and $|e_1,e_2\rangle$ is even and about $0.5$, see Fig.~\ref{fig:GSE}(b).
Related to the entanglement, also near-maximum coherence $|\langle g_1,g_2|\rho |e_1,e_2\rangle|\sim 0.5$ can be created via the dissipation.
%
%
The dashed lines in Fig.~\ref{fig:GSE} show the influence of pure dephasing. If the pure dephasing is low, then it only slightly reduces the fidelity of the achieved Bell state. For example, the entanglement for $2.8\leqslant|J|/\omega_0\leqslant 6.3$ decreases to about $0.87$ corresponding to a $95\%$ fidelity of the states $|\Phi_\pm\rangle$ if a pure dephasing of $\Gamma_\phi=0.01\gamma_0$ is taken into account. The coherence $|\langle g_1,g_2|\rho |e_1,e_2\rangle|$ is also only slightly reduced.

Next to the achievable fidelity, also the operation time is a crucial characteristic of any quantum gate. To explore the temporal dynamics of our system, we start in an initial product ground state  $|g_1,g_2\rangle$ with AOC set to zero. Then we switch on the AOC in a switching time $t_{sw}$. After a certain time, we switch off this coupling again.
A complication in the theoretical analysis arises from the fact that with the time-dependence of $J$, also our transformation to the dressed state basis becomes time-dependent. The bare basis is transformed to the dressed-state basis via an unitary operation defined as
\begin{equation}
 \mathfrak{T}=\begin{pmatrix}
 a  & 0     & 0      &b \\
 0  &\beta  &\alpha  &0 \\
 0  &\alpha &-\beta  &0 \\
 -b &0      &0       &a
\end{pmatrix}\,.
\end{equation}
Taking into account the time dependence of the coefficients introduces a coherent contribution \begin{align}
H_\mathfrak{T}=-i\hbar\mathfrak{T} \partial_t{\mathfrak{T}}^\dag 
\end{align}
to the transformation of Hamiltonian \cite{PRL101p253602}.
We thus obtain the  master equation
\begin{equation}
\label{timedep}
 \dot{\rho}=-\frac{i}{\hbar}\sum_j \omega_j[R_{jj}+H_\mathfrak{T},\rho]+\mathscr{L}_{SP}\rho+\mathscr{L}_{TP}\rho+\mathscr{L}_\phi\rho \,.
\end{equation}

\begin{figure}
 \centering
\includegraphics[width=0.9\columnwidth]{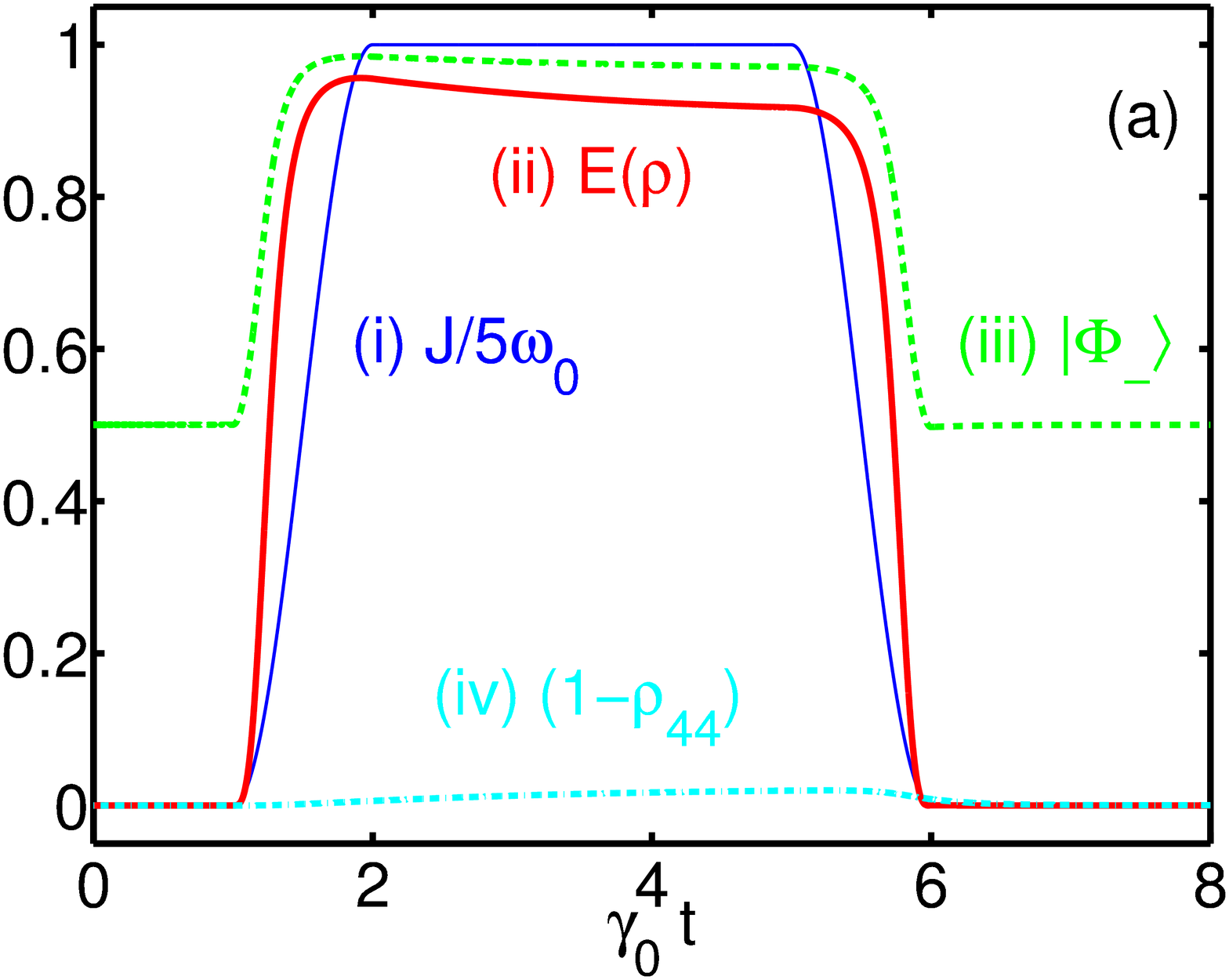}\\
\includegraphics[width=0.9\columnwidth]{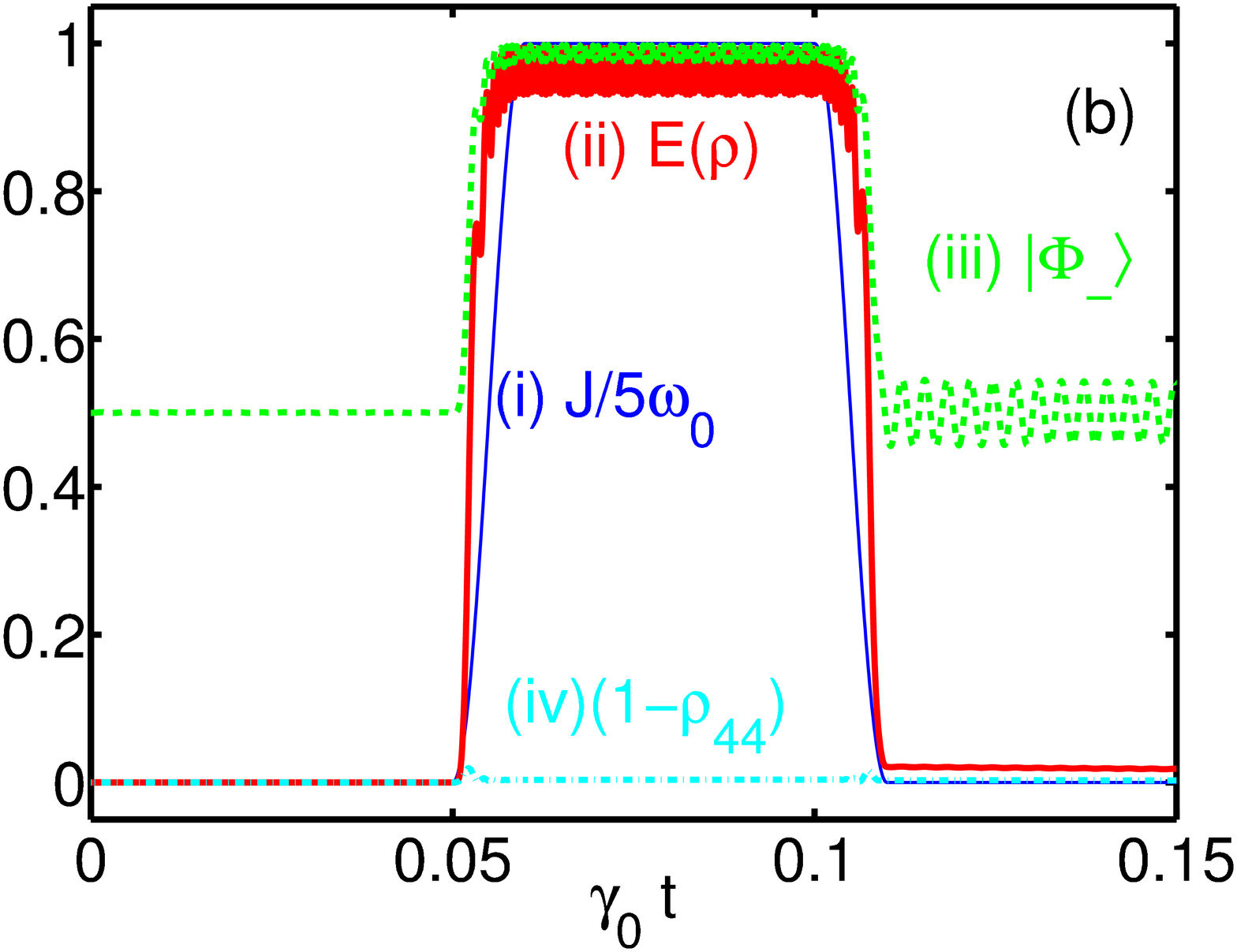}
\caption{\label{fig:GSEt}(Color online) Time-evolution of entanglement for time-dependent couplings between the qubits. The time-dependent coupling strength $J$ is shown as thin blue line (i). The amount of entanglement is shown as red line (ii). The fidelity of the  Bell state $|\Phi_-\rangle$ is indicated as the dashed green line (iii), and the total excitation from the ground state as dot-dashed cyan line (iv). Subpanel (a) shows the case of slow switching of the AOC $J$ with switching time $t_{sw}=\gamma_0^{-1}$. (b) shows the corresponding results with fast switching $t_{sw}=0.01\gamma_0^{-1}$ (b). The results are obtained for $r_{21}=\lambda_0/1000$ and include a pure dephasing of $\Gamma_\phi = 0.01\gamma_0$.}
\end{figure}

Results of the numerical solution of Eq.~(\ref{timedep}) are shown in Fig.~\ref{fig:GSEt}. Initially, the system is $|4\rangle$. The AOC is swept from a negligible value $\gamma_0$ to the large value $5\omega_0$ and then back to a vanishing coupling. In Fig.~\ref{fig:GSEt}(a), the switching of the AOC occurs rather slowly, with switching time on the order of the relaxation time,  $t_{sw}=\gamma_0^{-1}$. In this case, the system approximately remains in the time-dependent ground state throughout the whole evolution, with less than one percent of population in the three upper states during the switching. Due to the evolution of the ground state, the system changes from a product state $|g_1,g_2\rangle$ to a highly entangled state $|\Phi_-\rangle$ with fidelity of $0.98$ and corresponding entanglement $E(\rho)=0.95$, and back to the product state again when we turn off the AOC.

Next, we assume a faster switching time, see Fig.~\ref{fig:GSEt}(b). Although the switching time is only  $t_{sw}=0.01\gamma_0^{-1}$, still more than $98\%$ population remain in the ground state $|4\rangle$ at all times. The fidelity and entanglement are similar to the case of slow switching. Thus, a fast and robust entangling operation can be realized in our system with the use of a tunable AOC. Both highly entangled states $|\Phi_\pm\rangle$ can be prepared from the product state $|g_1,g_2\rangle$ via tuning on a large AOC, without any external coherent driving fields. During this operation, the system remains in its ground state. Only non-adiabatic effects during faster switchings of the AOC give rise to a partial excitation of the upper states. Thus, this operation is nearly free from decoherence.

\begin{figure}[t]
\centering
\includegraphics[width=0.9\columnwidth]{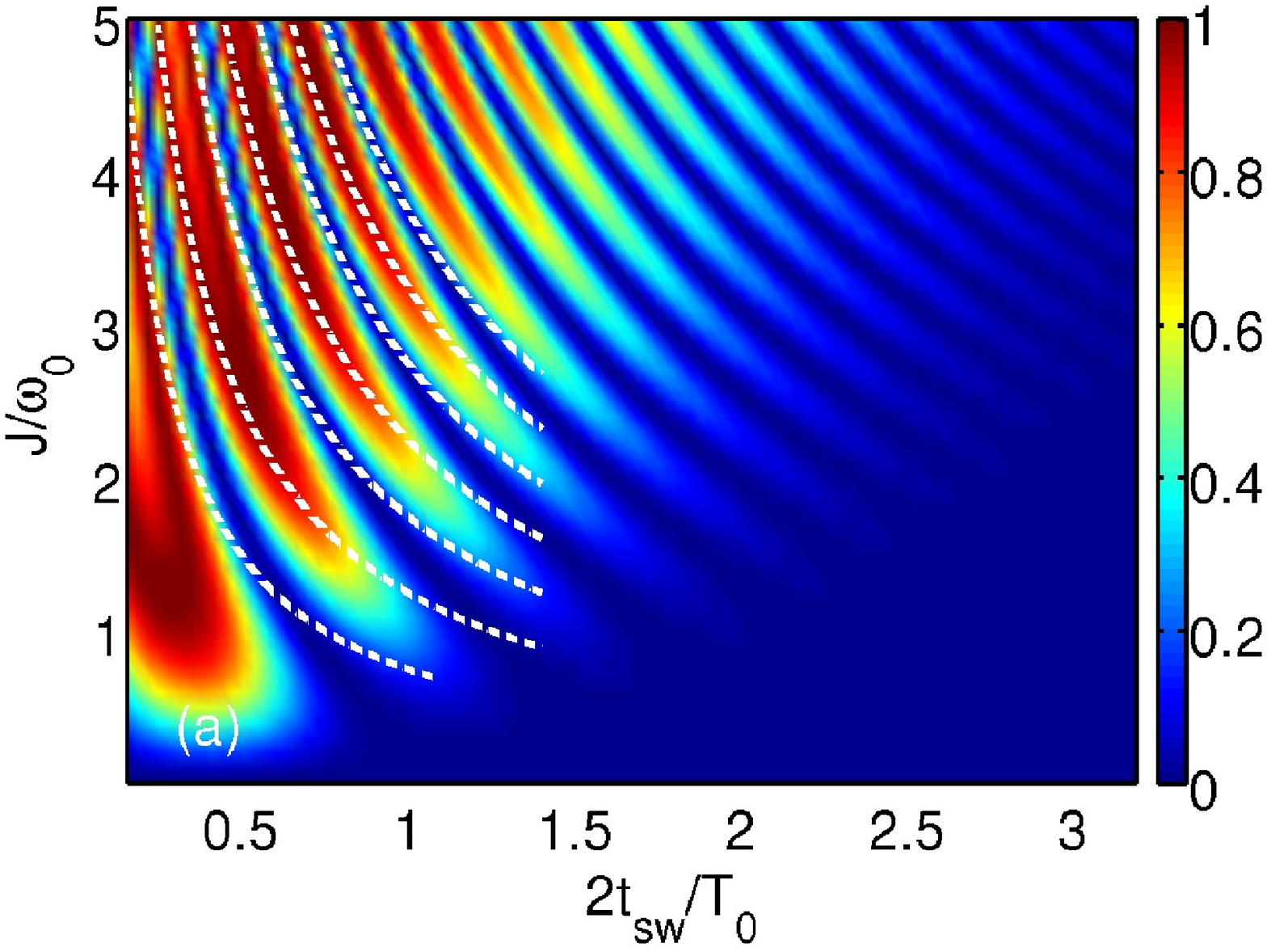}\\
\includegraphics[width=0.9\columnwidth]{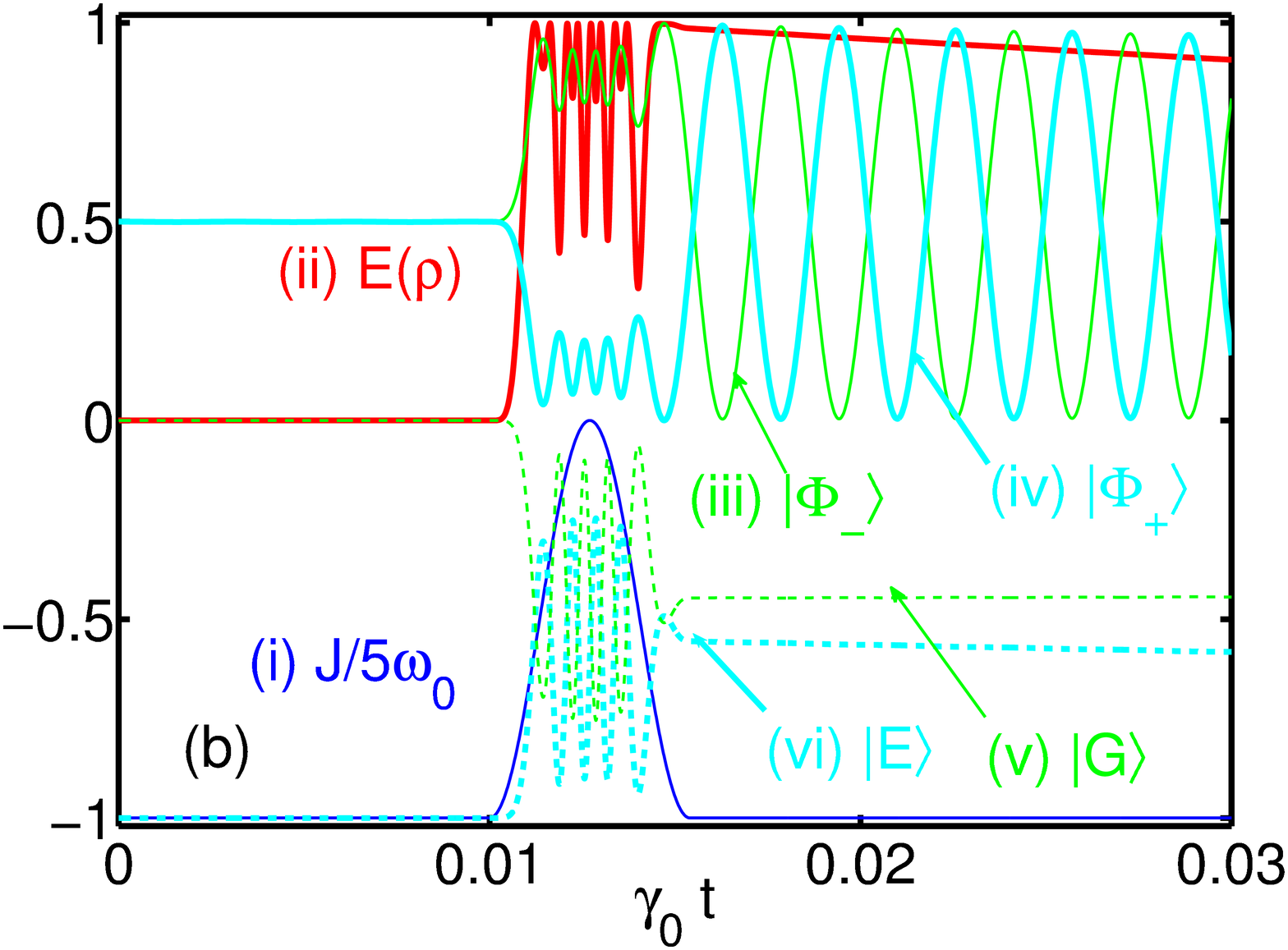}
\caption{\label{fig:GSEtKEnt}(Color online) (a) Amount of entanglement as a function of the switching time $t_{sw}$ and the coupling $J$. The white dashed lines indicate pulses with area $3\pi/2,5\pi/2,7\pi/2,9\pi/2,11\pi/2,13\pi/2,15\pi/2$, respectively.
(b) shows the time evolution of  the entanglement generation for a pulsed AOC $J$. The time evolution of the coupling $J$ is shown as blue line (i). The resulting entanglement is plotted as red line (ii). The fidelity of the state $|\Phi_-\rangle$ is shown as green line (iii); that of the state $|\Phi_+\rangle$ as cyan line (iv). Finally, the state population of 
$|G\rangle$ is shown as green dashed line (v), and the population of $|E\rangle$  as cyan dashed line (vi). Note that the lines (i), (v) and (vi) are shifted down by $-1$ for better clarity. In (b), the switching time is $t_{sw}=0.0027\gamma_0^{-1}$.
The other parameters are: $\Gamma_\phi=0.01\gamma_0$ and $r_{21}=\lambda_0/1000$.} 
\end{figure}

Interestingly, in Fig.~\ref{fig:GSEt}(b) it can be seen that the fidelity of $|\Phi_-\rangle$ oscillates rapidly after the switching sequence is finished. These oscillations at constant AOC arise from the slight excitation of upper states due to the fast switching of a large AOC. This partial  excitation induces coherences which oscillate at the transition frequency. The excitation $\rho_E$ of state $|E\rangle$ then gives rise to an oscillation of fidelity of $|\Phi_-\rangle$ with amplitude $\sqrt{\rho_E}$. For example, the excitation $\rho_E$ is about $0.0025$ after the AOC is switched off  at $\gamma_0 t =0.11$. Correspondingly, the  fidelity of $|\Phi_-\rangle$ oscillates with amplitude $0.05=\sqrt{0.0025}$ around the mean value $0.5$. 

 \begin{figure}[t]
  \centering
  \includegraphics[width=0.9\columnwidth]{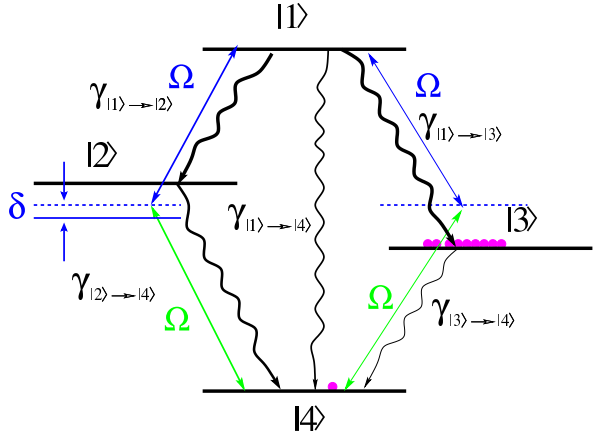}
  \caption{\label{fig:OneFldTrpSetup}(Color online) Level diagram and relevant coherent and incoherent processes for the generation of stationary state entanglement with a single continuous wave driving field. As an example, we use the parameters correspond to those indicated by the point P in Fig.~\ref{fig:OneFldTrpEnt}(a).}
 \end{figure}

We finally turn to shortest switching times of order of the inverse transition frequency of  the qubit. Since the AOC $J$ exceeds the transition frequency $2 \omega_0$ of the doubly excited state $|E\rangle$ at its maximum value, it should in principle be capable of exciting the qubit system to the excited state to a large degree. Similarly, due to non-adiabatic effects, large residual entanglement after the switching should be possible. This is demonstrated in Fig.~\ref{fig:GSEtKEnt}(a), which shows the entanglement between the two qubits in dependence on  both the AOC strength $J$ and the switching time $t_{sw}$. The duration of the AOC is $2t_{sw}$, and $T_0=2\pi/\omega_0$ is the carrier period. We find that the entanglement of the system can be large after the AOC turn off if the area of the AOC pulse satisfies
\begin{align}
2\Omega_{Rabi}t_{sw}=n\pi/2\,,
\end{align}
where $n\in\{3, 5, 7,9,\ldots\}$, as indicated by the dashed lines in Fig.~\ref{fig:GSEtKEnt}(a). Here, the effective Rabi frequency is given by 
\begin{align}
\Omega_{Rabi}=\omega_0\sqrt{\left(\frac{J}{\omega_0}\right)^2+4\left(\frac{2\pi}{2t_{sw}\omega_0}-1\right)^2}\,.
\end{align}
The high residual entanglement goes along with a high population of the excited state $|E\rangle$. 
Efficient population of  $|E\rangle$ can be expected for $2t_{sw}/T_0 \sim 1$. As shown in Fig.~\ref{fig:GSEtKEnt}(b), the states $|E\rangle$ and $|G\rangle$ are then evenly populated after the switching operation is finished. Interestingly, the entanglement of the system can be $E(\rho)=0.98$ after an operation with $J\sim 1.5\omega_0$ and $2t_{sw}\sim T_0/4$. After the switching, the population oscillates between the two Bell states $|\Phi_\pm\rangle$, as they are not the eigenstates of the system. The entanglement itself, however, does not show this oscillation. 

We thus have demonstrated an experimentally realizable mechanism for the generation of entanglement and for generating maximum coherence in time scales of order of the inverse qubit transition frequency, exclusively driven by dissipation. In contrast to intuition, the dissipation turns into a favorable resource if combined with the strong coupling between the two qubits. It can be seen from Fig.~\ref{fig:GSE} that the ground state is continuously tunable with the restriction that the population of the excited state $|E\rangle$ remains smaller than 50\%. 

\begin{figure}[t]
 \centering
 \includegraphics[width=0.9\columnwidth]{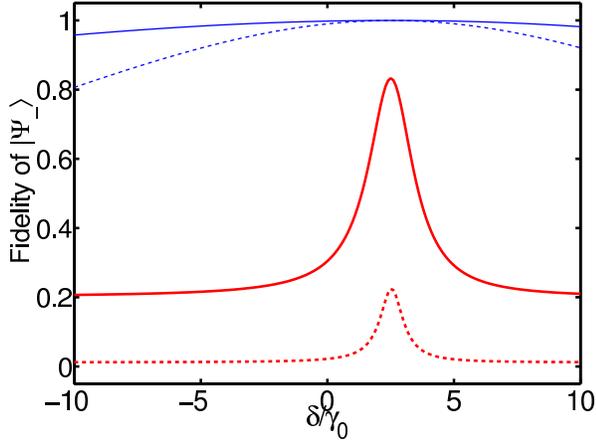}
 \caption{\label{fig:OptDtn}(Color online) Optimizing the entanglement generation with a single cw field. The figure shows the fidelity of the Bell state $|\Psi_-\rangle$ as a function of the detuning $\delta$. The solid lines are drawn for a distance $r_{21}=\lambda_0/50$, the dashed lines show a smaller separation $r_{21}=\lambda_0/1000$. The two upper blue lines correspond to the ideal system without pure dephasing, whereas two lower red lines shows the results taking into account a pure dephasing $\Gamma_\phi=0.01\gamma_0$.}
\end{figure}
\subsection{Excited state entanglement}
In the previous section, we have considered entanglement generation in the ground state. Here, we analyze the generation of entanglement between two excited qubit states. Usually, such excited states are subject to dissipation, and thus have only a limited life time. Therefore, in the spirit of decoherence free subspaces~\cite{PRA60p1944,PRL92p147901}, we will analyze the possibility of generating trapped entangled excited states which are immune to the dissipation of system, and focus on the states
\begin{align}
\label{exc}
|\Psi_\pm \rangle=\frac{1}{\sqrt{2}}(|e_1,g_2\rangle\pm |g_1,e_2\rangle)\,.
\end{align}

In atomic systems with weak coupling, it is challenging to highly populate the states in Eq.~(\ref{exc}). The anti-symmetric state is decoupled from the driving field dynamics if both atoms are subject to the same driving field with the same phase, as it is usually the case for two atoms with a distance of order of the transition wavelength~\cite{PR372p369}. On the other hand, it is also difficult to robustly create the symmetric state because it couples to both the collective  ground and excited states~\cite{ZFicekQuantInterfCoh2004}.

\begin{figure}[t]
 \centering
 \includegraphics[width=0.9\columnwidth]{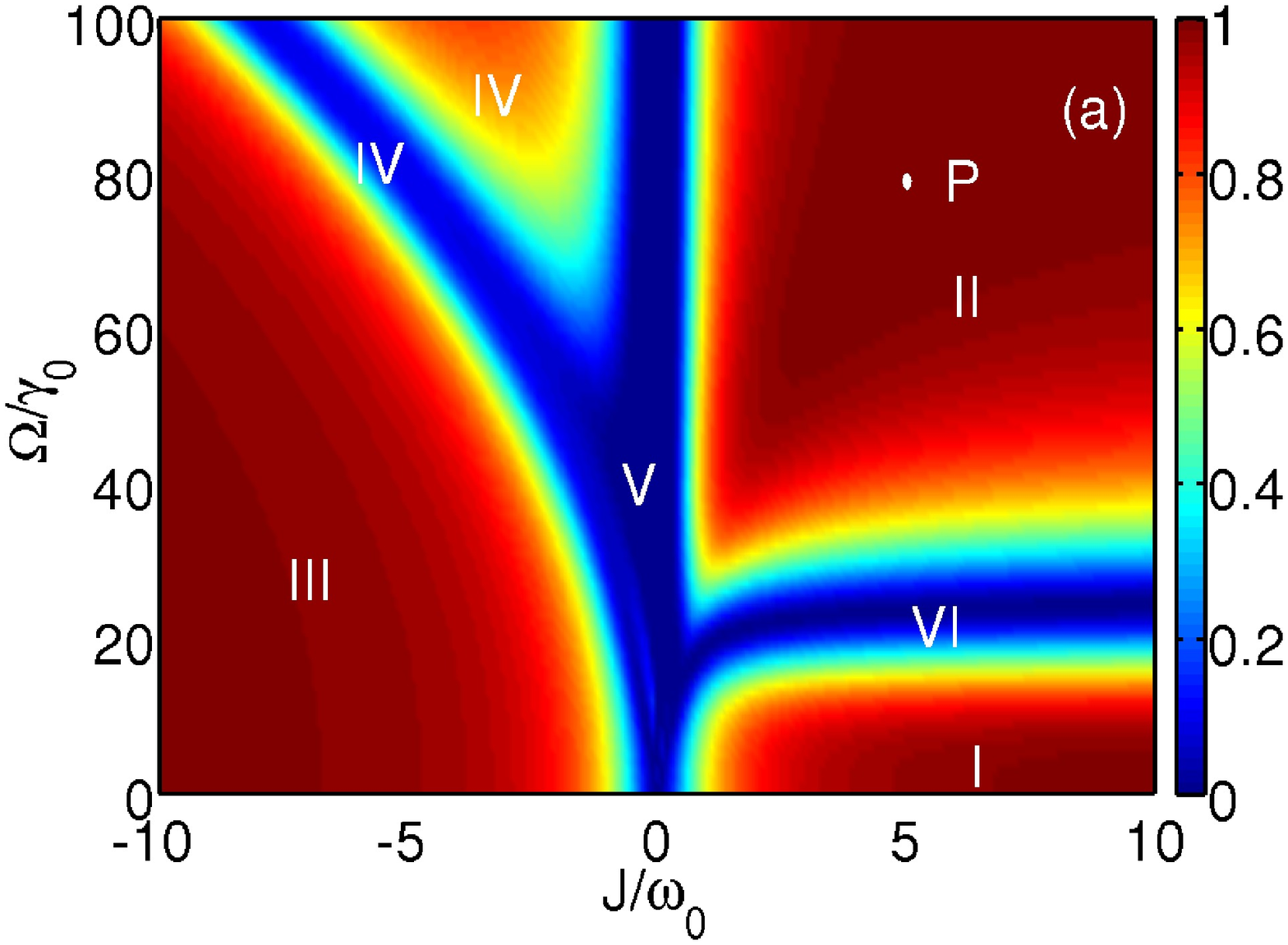}\\
\vskip 0.1cm
\includegraphics[width=0.9\columnwidth]{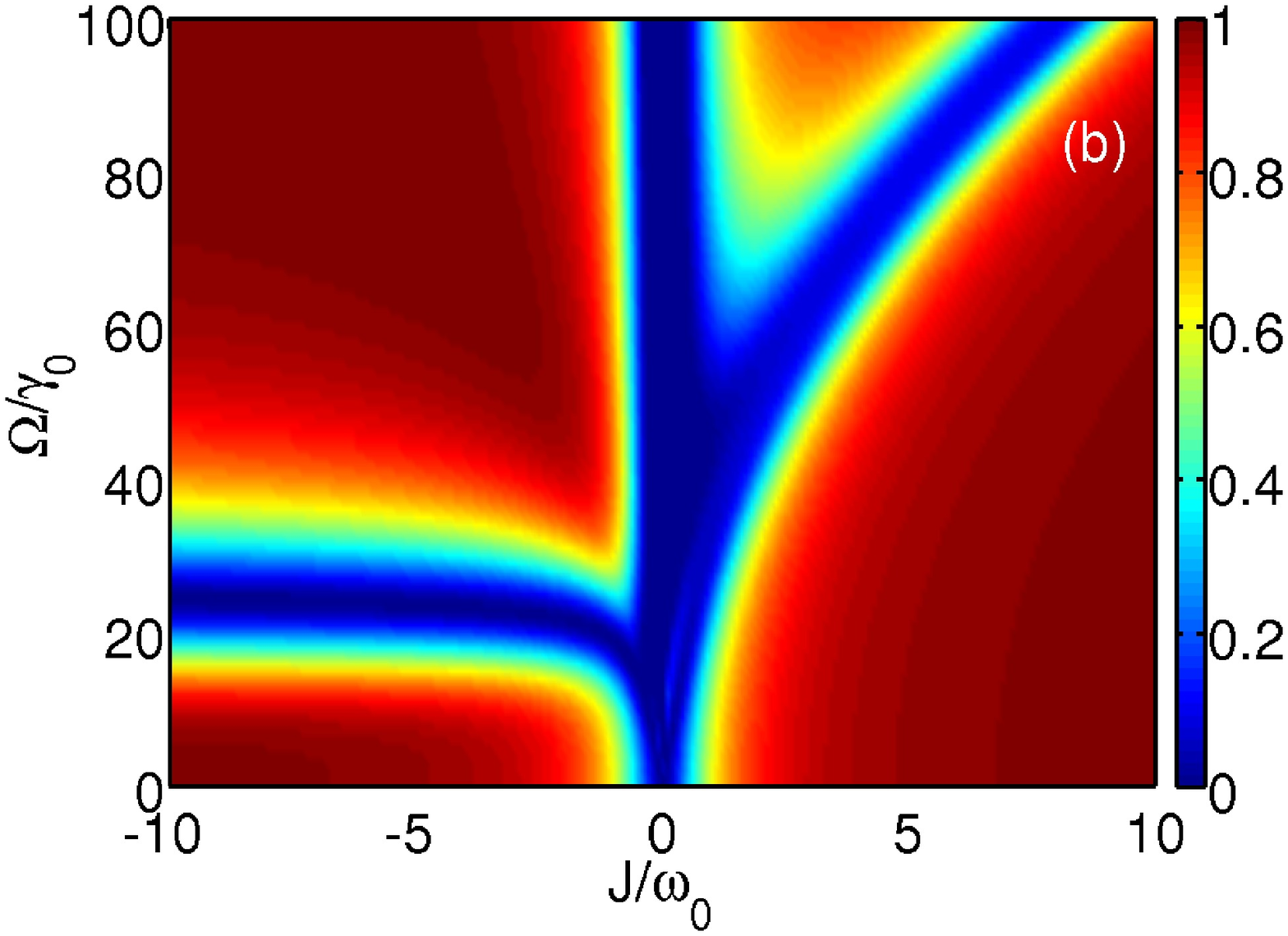}
 \caption{\label{fig:OneFldTrpEnt}(Color online) (a) Steady-state entanglement with a single continuous-wave driving field as a function of the driving field Rabi frequency $\Omega$ and the coupling $J$. The upper panel (a) shows the results for parallel dipole moments, $p=1$, of the two qubits. The lower panel (b) shows the corresponding results for anti-parallel dipole moments,  $p=-1$. In both cases, the frequency of the driving field is chosen as half the transition transition frequency of the $|1\rangle \rightarrow |4\rangle$ transition. Thus, for large coupling $J$, all transitions are far off-resonant with the field. The phase is chosen as  $\phi_{12}=0$.}
\end{figure}

Superconducting circuits differ from atomic systems in many ways, and these differences can be exploited to populate the states Eq.~(\ref{exc}). Methods to robustly prepare  the antisymmetric state have been discussed in our previous work \cite{PRB79p024519} using adiabatic passage. For this, first a difference between the two transition frequencies of the qubits is introduced during the population stage, and then this difference is tuned to zero to protect the prepared state. Accordingly, the generated antisymmetric state decays at a very small rate. Ojanen et. al \cite{PRB76p100505R} have also suggested a superconducting circuit to prepare the antisymmetric state. They designed the circuit geometrically such that a microwave $\pi$ pulse inserted in  a coplanar superconducting cavity anti-symmetrically drives the two coupled qubits to their  antisymmetric state. However, in this setup, also the environmental noise in the bath (cavity) couples to the antisymmetric state. As a result, the antisymmetric state will decay fast. Furthermore, the geometry is fixed and the transitions involving the symmetric state are forbidden. Thus it is difficult to highly populate the symmetric state.

Motivated by this, in the following, we propose two methods to prepare and trap the entangled anti-symmetric and symmetric states.

\begin{figure}[t]
 \centering
\includegraphics[width=0.45\columnwidth]{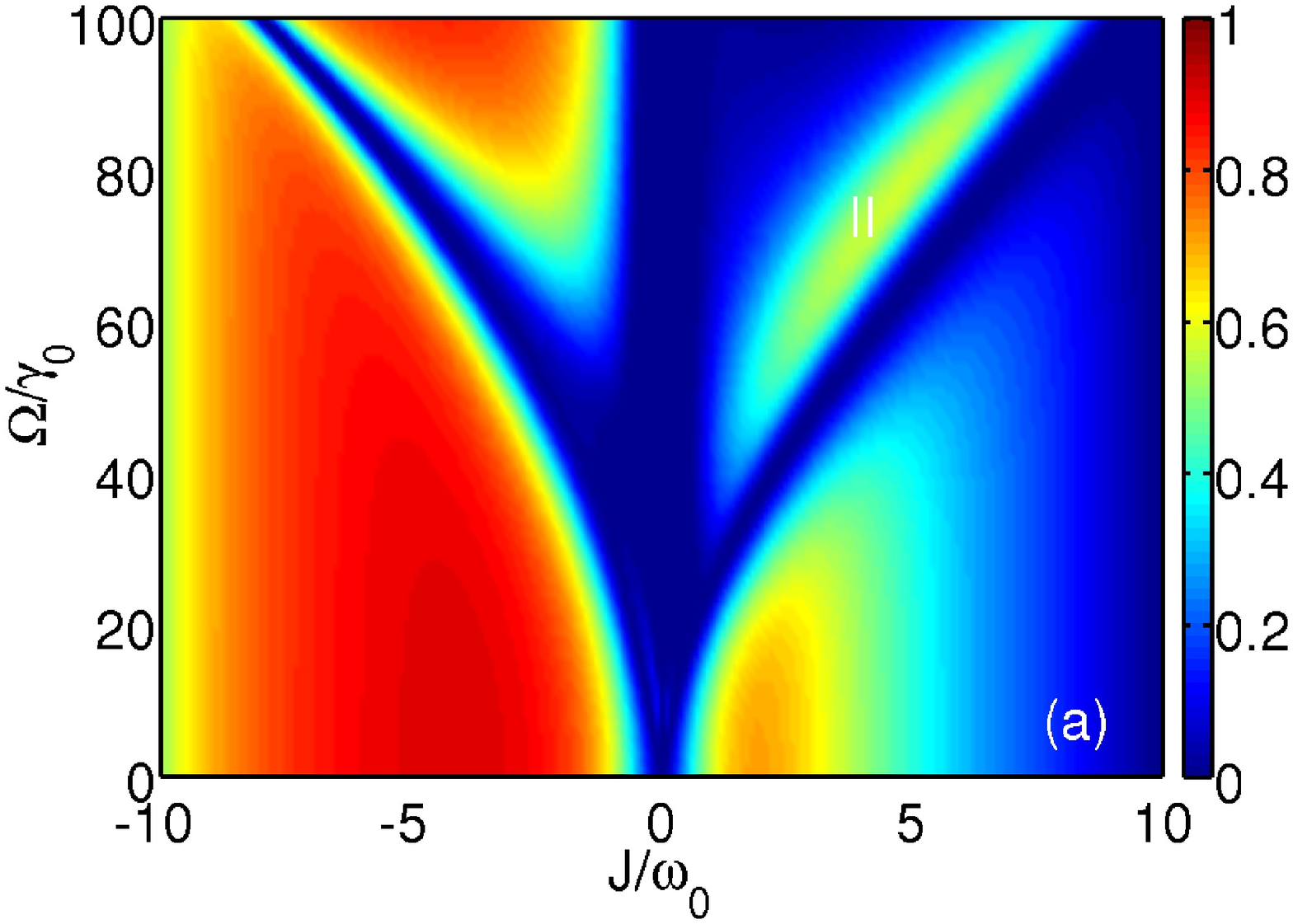}
\includegraphics[width=0.45\columnwidth]{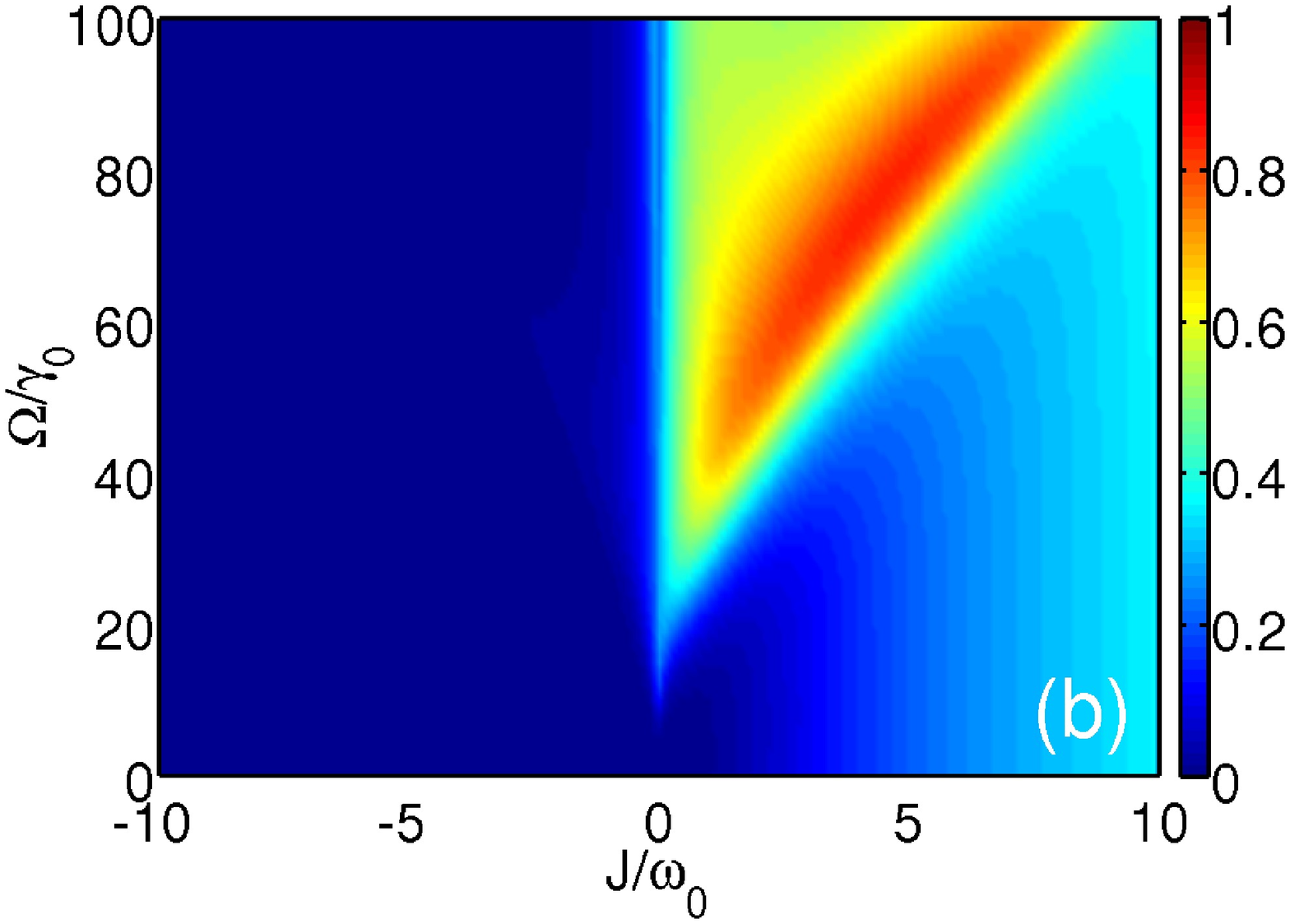}\\
\includegraphics[width=0.45\columnwidth]{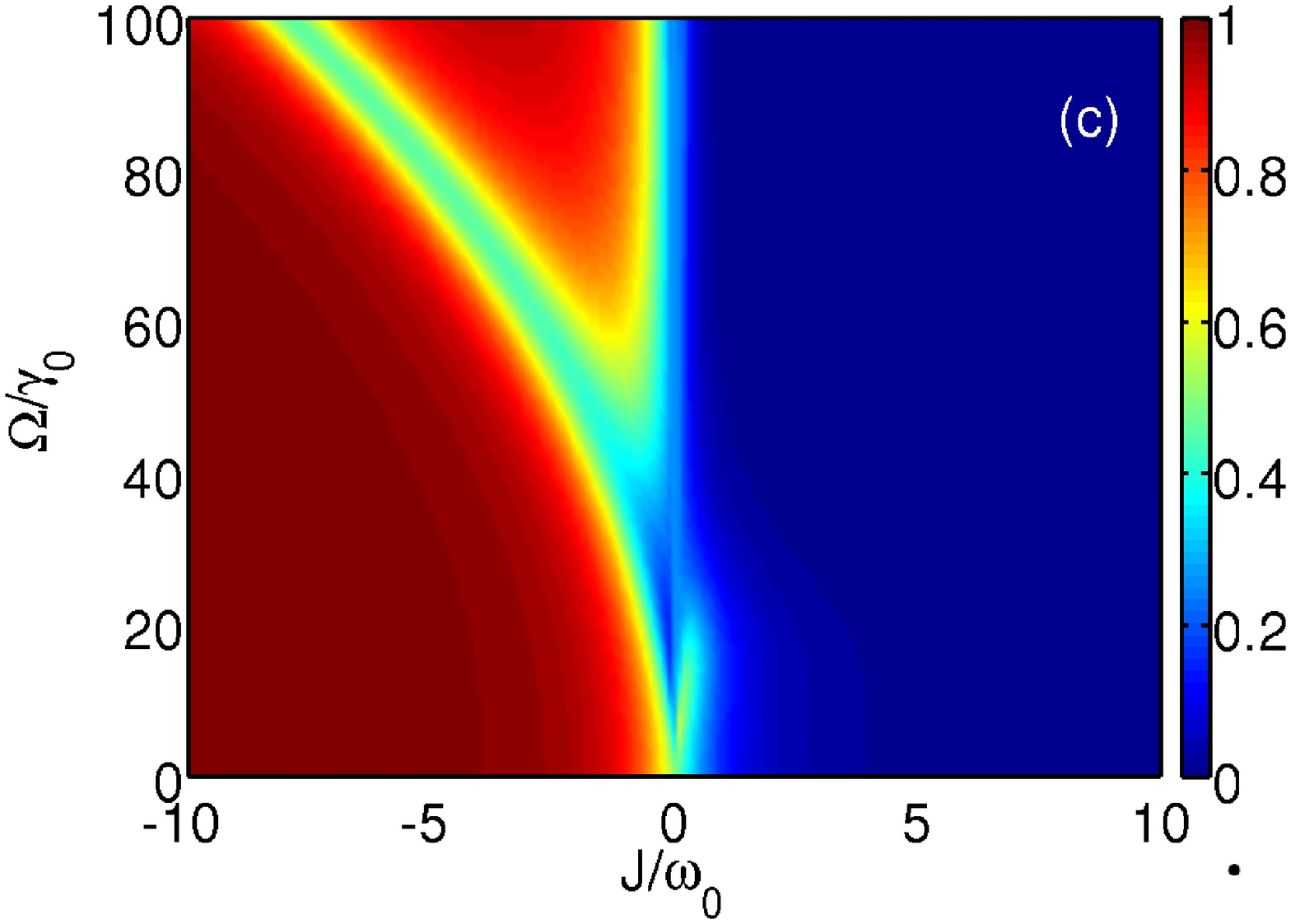}
\includegraphics[width=0.45\columnwidth]{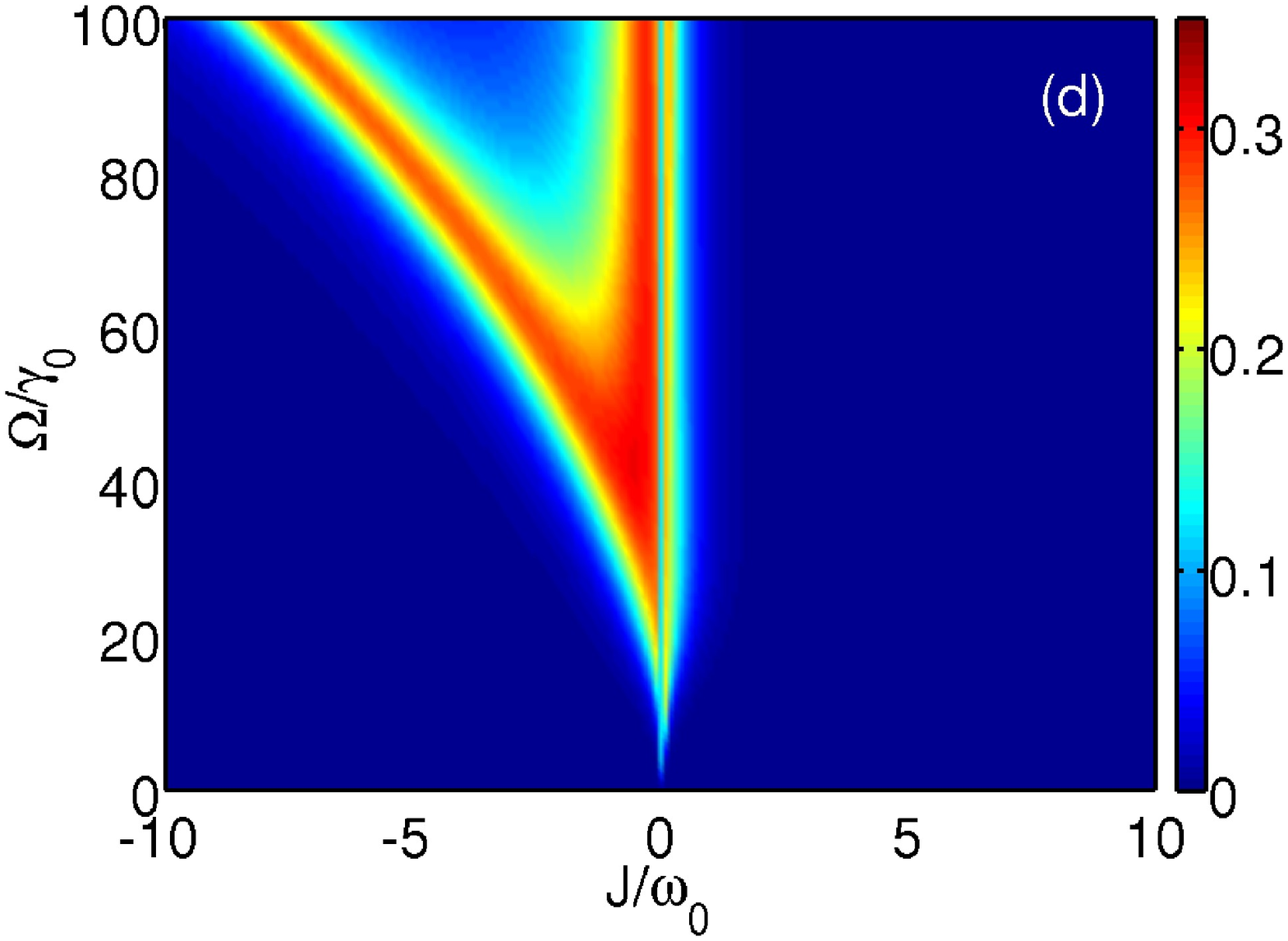}
\caption{\label{fig:OneFldBllF}(Color online) Interpretation of the steady state entanglement shown in Fig.~\ref{fig:OneFldTrpEnt} in terms of the population of the magic state basis states. The four sub-panels show the fidelity of the states (a) $|\Phi_-\rangle$, (b) $|\Psi_-\rangle$, (c) $|\Phi_+\rangle$, and (d) $|\Psi_+\rangle$. The figure is drawn for parallel dipole moments ($p=1$).}
\end{figure}

%
\subsection{Trapping assisted by a single field}

Here, we discuss the creation of stationary entanglement by using only one continuous-wave driving field. As both states $|\Psi_\pm \rangle$ are maximally entangled, the system acquires large stationary entanglement if either of these two states is trapped with high fidelity.

The considered configuration is shown in Fig.~\ref{fig:OneFldTrpSetup}. The system is driven by one cw field with Rabi frequency $\Omega$. In the interaction picture defined by the unitary transformation
\begin{subequations}
\label{eq:U}
\begin{align}
 U &=\sum_{j=1}^4 e^{i\omega_{Rj}t|j\rangle \langle j|} \,,\\
 \omega_{R1} &=2\omega_L+\omega_4 \,,\\
 \omega_{R2} &=\omega_L+\omega_4 \,,\\
 \omega_{R3} &=\omega_L+\omega_4 \,,\\
 \omega_{R4} &=\omega_4 \,,
\end{align}
\end{subequations}
the master equation for the system density matrix $\rho$ takes the form
\begin{equation}\label{eq:MEQ1}
 \dot{\rho}=\frac{i}{\hbar}[\tilde{H}_0+\tilde{H}_I+H_{BIE},\rho]+\mathscr{L}_{SP}\rho+\mathscr{L}_{TP}\rho\,,
\end{equation}
where $\tilde{H}_0=\sum_j \Delta_j |j\rangle \langle j|$ is the free-evolution Hamiltonian. The interaction Hamiltonian is
\begin{align}
 \tilde{H}_I&=\hbar \Omega\left\{a\left(\alpha+\beta p e^{-i\phi_{21}} \right)+b\left(\beta+\alpha p e^{-i\phi_{21}} \right)\right\}R_{12}\,\nonumber\\\nonumber 
 &+\hbar\Omega\left\{a\left(\alpha-\beta p e^{-i\phi_{21}} \right)+b\left(\beta-\alpha p  e^{-i\phi_{21}} \right)\right\}R_{34}\,\\ \nonumber
  &+\hbar\Omega\left\{a\left(\beta+\alpha p e^{-i\phi_{21}}\right)-b\left(\alpha+\beta p e^{-i\phi_{21}} \right)\right\}R_{24}\,\\ 
&-\hbar\Omega\left\{a\left(\beta-\alpha p e^{-i\phi_{21}} \right)-b\left(\alpha-\beta p e^{-i\phi_{21}} \right)\right\}R_{13}+ \textrm{H.c.}\,,
\end{align}
and the detuning are defined as
\begin{subequations}
\begin{align}
  \Delta_1 &=2\delta \,,\\
  \Delta_2 &=\delta+\omega_2-\frac{\omega_1+\omega_4}{2} \,,\\
  \Delta_3 &=\delta+\omega_3-\frac{\omega_1+\omega_4}{2} \,,\\
  \Delta_4 &=0 \,,
\end{align}
\end{subequations}
where $\delta=(\omega_1-\omega_4)/2-\omega_L$. The detuning $\delta$ can be controlled via the frequency of the applied TDMF. The phase $\phi_{21}=k_Lr_{21}$ of the driving field depends on the frequency of the driving field and is non-zero. Therefore, the channel relevant to $|3\rangle$ needs to be included. Note that the coherent part $H_{BIE}$ resulting from DDI remains unchanged under the applied transformation.

To analyze the system performance, we choose a driving field with Rabi frequency $\Omega=80\gamma_0$, and an AOC $|J|=5\omega_0$. Sweeping the frequency of the driving field, we find an optimal detuning $\delta_{opt}=2.5\gamma_0$ leading to maximum fidelity in the target state, see Fig.~\ref{fig:OptDtn}.
In the absence of pure dephasing, one can create the Bell state $|\Psi_-\rangle$ with a high fidelity in a broad frequency window for both considered separations $r_{21}=\lambda_0/50$ or $r_{21}=\lambda_0/1000$. At optimal detuning $\delta=2.5\gamma_0$, the fidelity is one. 
Including pure dephasing with $\Gamma_\phi=0.01\gamma_0$, the fidelity reduces to a small value of about $0.22$ for small separation. In contrast, one can still trap the state $|\Psi_-\rangle$ with a fidelity of $0.83$ for larger distances $r_{21}=\lambda_0/50$.
The reason is that in the case of small separation, the decay from $|1\rangle$ to $|3\rangle$ and thus to $|\Psi_-\rangle$ is several orders of magnitude smaller than that to $|2\rangle$, which decays quickly to the ground state $|4\rangle$. But for $r_{21}=\lambda_0/50$, the decay rate $\gamma_{|1\rangle \rightarrow  |3\rangle}$ is comparable to $\gamma_{|1\rangle \rightarrow  |2\rangle}$ and several orders of magnitude larger than $\gamma_{|3\rangle \rightarrow  |4\rangle}$. Therefore, most population decays to state $|3\rangle$ for larger separation.
Similarly, one can obtain the comparable trapping in the state $|\Psi_+\rangle$ at $\delta=2.5\gamma_0$ if $J=-5\omega_0$.

Figure~\ref{fig:OneFldTrpEnt} shows the entanglement of the system against the driving field Rabi frequency and the AOC strength in the absence of pure dephasing. The system is highly entangled in three regions (I, II, III) in either case of parallel ($p=1$) or anti-parallel ($p=-1$) dipole moments. In these regions the AOC strength $J$ is larger than the transition frequencies. In  region IV, the system is weakly entangled, and in regions V, VI and VII, it is almost disentangled.

\begin{figure}[t]
 \centering
 \includegraphics[width=0.9\columnwidth]{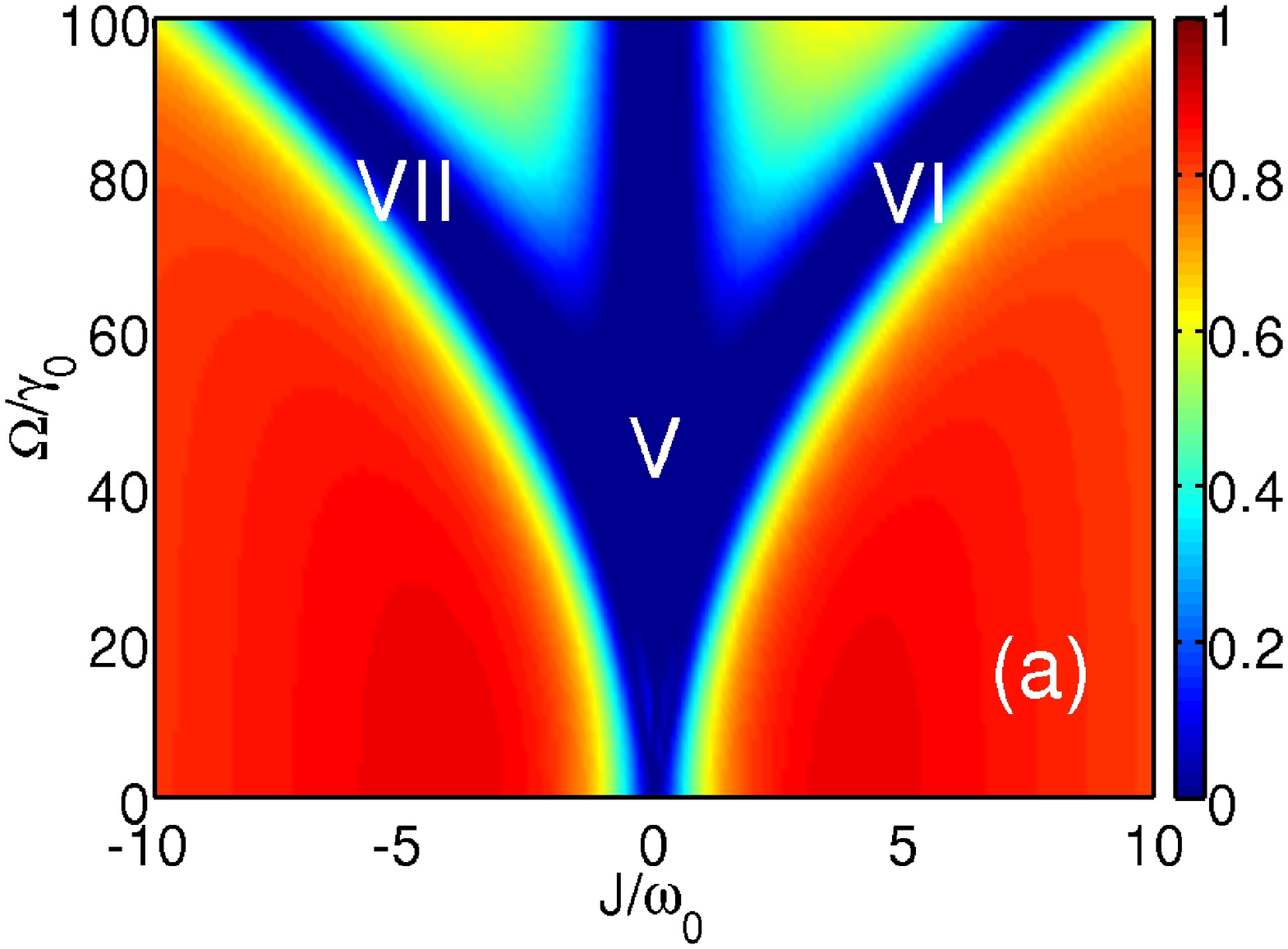}\\
 \includegraphics[width=0.9\columnwidth]{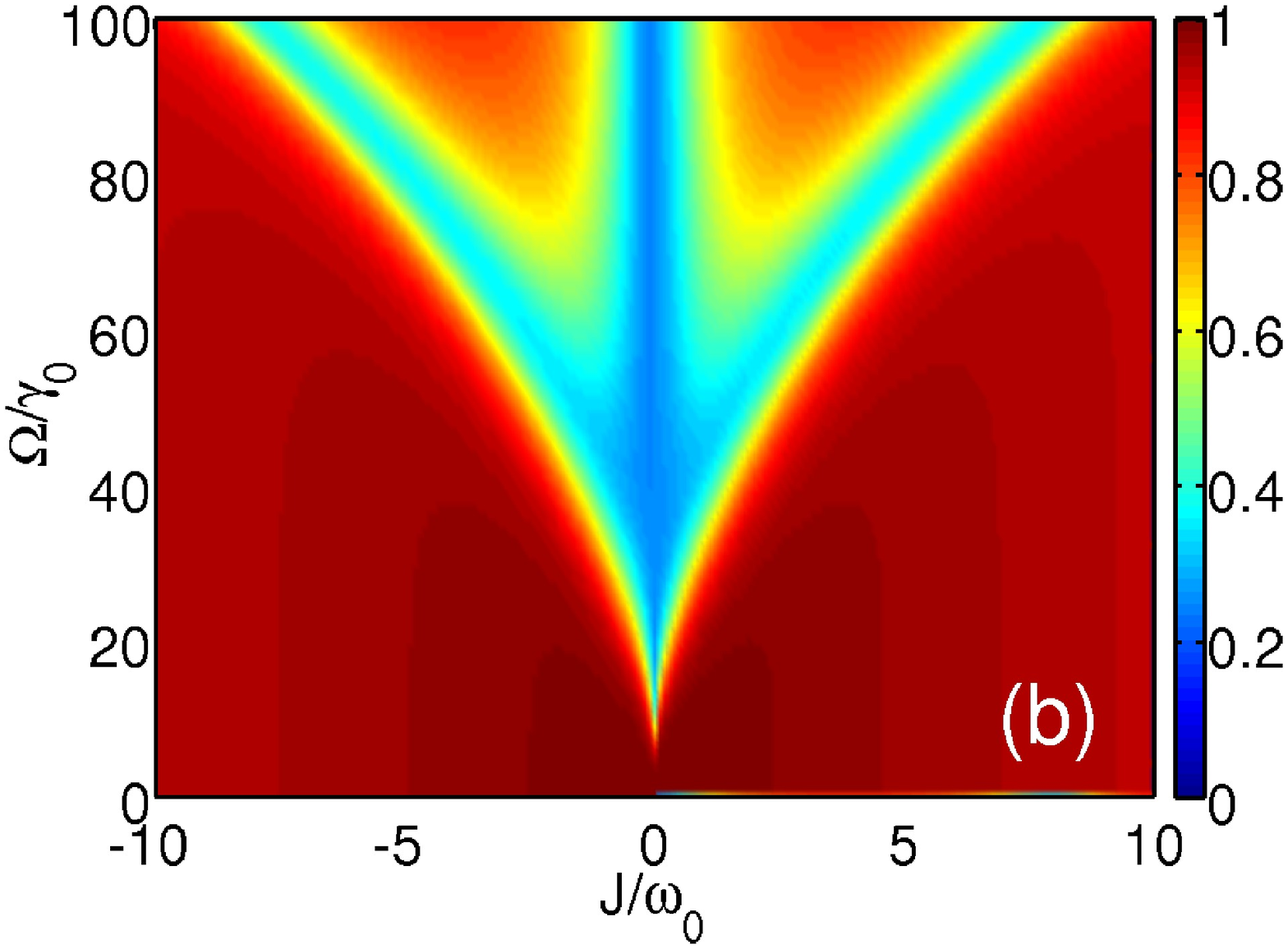}\\
 \includegraphics[width=0.9\columnwidth]{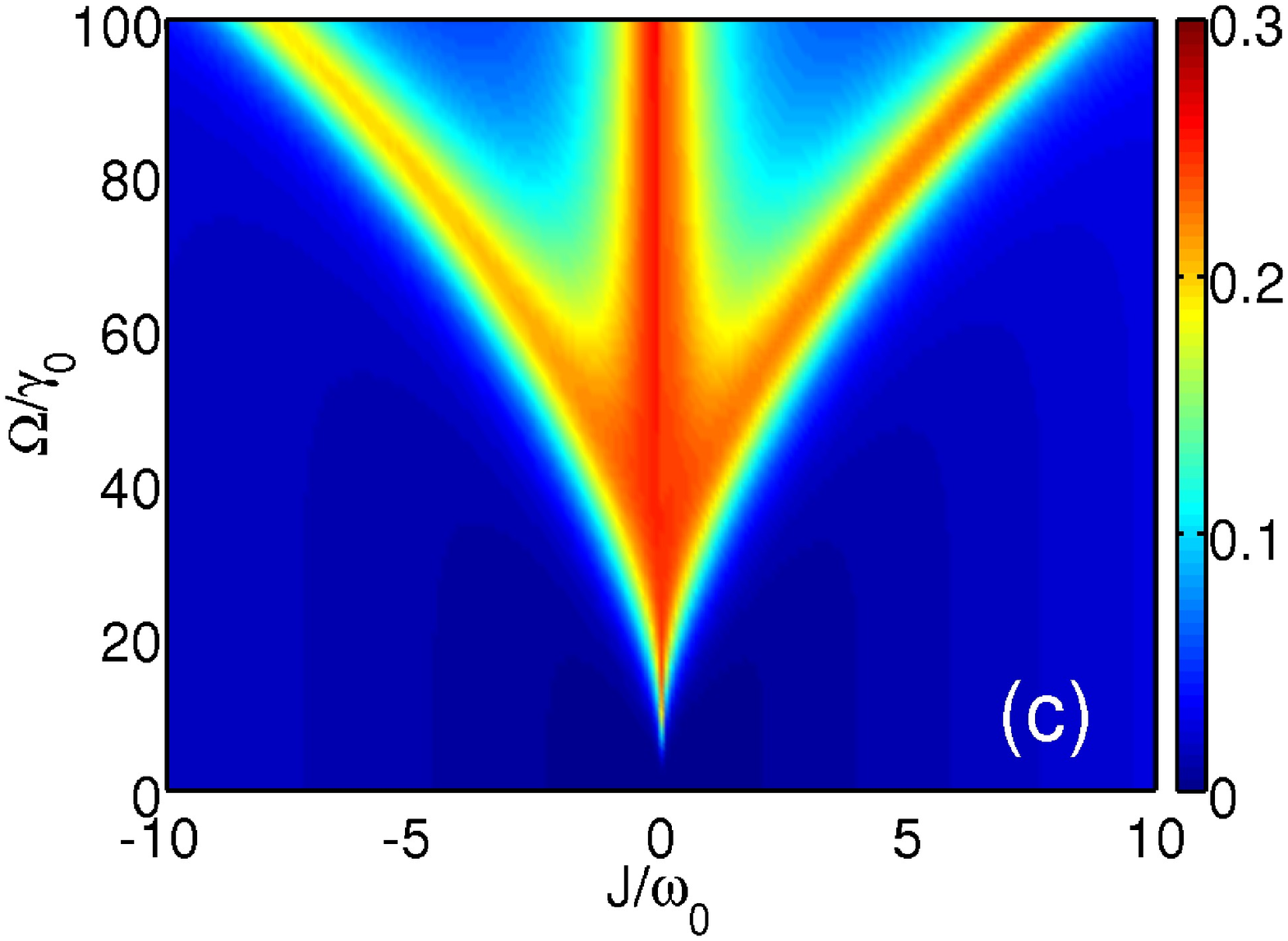}
 \caption{\label{fig:OneFldTrpPDph}(Color online) Steady-state entanglement and interpretation in terms of Bell state populations against the driving field Rabi frequency $\Omega$ and the coupling $J$. The figure is similar to Fig.~\ref{fig:OneFldTrpEnt}, but includes pure dephasing $\Gamma_\phi=0.01\gamma_0$. (a) shows the amount of entanglement, (b) the fidelity of the Bell state $|\Phi_-\rangle$, and (c) that of the Bell state $|\Psi_-\rangle$. The results are shown for anti-parallel dipole moments ($p=-1$).}
\end{figure}

The structure of entanglement shown in Fig.~\ref{fig:OneFldTrpEnt}(a) can be understood as follows. We decompose the system state into the the magic state basis \cite{PRL78p5022}, show the fidelity of the four magic states in Fig.~\ref{fig:OneFldBllF}, and discuss the different regions separately in the following. 
In region I, the driving field is weak, and off-resonant from all transitions.  As a result, almost all population is in the ground state, which is close to the Bell state $|\Phi_-\rangle$ for the considered large AOC $J$. This expectation is confirmed in Fig.~\ref{fig:OneFldBllF}(a).
From Fig.~\ref{fig:OneFldBllF}(b) we see that interestingly, the antisymmetric state can be prepared using a single driving field in region II. The field first pumps the system into the intermediate state $|2\rangle$ and then into the excited state $|1\rangle$, which subsequently decays to the ground state via three channels: the symmetric, the antisymmetric and the two-photon channel. Essentially different from the weak AOC case, the decay from the antisymmetric state to the ground state is strongly modified to be much smaller than other channels by the strong AOC, as shown in Fig.~\ref{fig:relJ}. Thus effectively the system is pumped into the antisymmetric state and remains trapped in it. This leads to high entanglement. For example, $E>0.95$ is achieved for $\Omega_1=\Omega_2=80\gamma_0$ and $J=5\omega_0$, see Fig.~\ref{fig:OneFldTrpEnt}(a). It should be noted that this state is created in the steady state of system. Therefore it is independent of the initial state protected from the relaxation, thus forming a decoherence-free subspace.
For negative $J$, in region III,  almost all population decays to the ground state which is now close to the Bell state $|\Phi_+\rangle$, see \ref{fig:OneFldBllF}(c). The reason is that the decay from $|1\rangle$ to the intermediate states $|3\rangle$ and those of these intermediate states are equal but much larger than that to $|2\rangle$, see Fig.~\ref{fig:relJ}. Thus only a small amount of population resides in the symmetric state $|3\rangle$ which means that the fidelity of the symmetric state is small. Again, the system is highly entangled in steady state.
At certain driving fields, e.g., in the region IV, the upper states $|2\rangle$ and $|3\rangle$ are partly excited. The system then dominantly is in a superposition of $|\Phi_+\rangle$ and $|\Psi_+\rangle$ with equal weight. This results in an unentangled state. Similarly, the system is unentangled in region VI.
In region V, the AOC is small. This reproduces the regular case of two weakly coupled two-level systems. The system is weakly excited by the off-resonant driving field, and the population of all four states are similar, see \ref{fig:OneFldBllF}. Thus the system is unentangled.

Further investigation shows similar entanglement properties as a function of the AOC and driving field for opposite dipole moments $\vec{d}_1 \upharpoonleft \downharpoonright \vec{d}_2 $, see Fig.~\ref{fig:OneFldTrpEnt}(b). In this case, the symmetric state $|\Psi_+\rangle$ can be trapped by a strong driving field in the region of negative coupling.

If a pure dephasing of $\Gamma_\phi=0.01\gamma_0$ is included, again the entanglement properties shown in Fig.~\ref{fig:OneFldTrpPDph}(a)  depend on the effective distance of the qubits. For small separation such as $r_{21}=\lambda_0/1000$, the system  is entangled only if almost all population is in its ground state, see Fig.~\ref{fig:OneFldTrpPDph} (b). If the AOC is small, populations in all three upper states are nearly equal and have similar structures as the state $|\Psi_-\rangle$ shown in Fig.~\ref{fig:OneFldTrpPDph}(c). Their maxima all are about $0.26$. As a result, the entanglement is nearly zero in the regions marked as $V,VI,VII$ in Fig.~\ref{fig:OneFldTrpPDph}(a).
For small separations, either the decay rate $\gamma_{|1\rangle\rightarrow |3\rangle}$ or $\gamma_{|1\rangle\rightarrow |2\rangle}$ is orders of magnitude smaller than $\gamma_0$, see Fig.~\ref{fig:relJsmallR}. The system thus reduces to a ladder-type three level system. In addition, the incoherent exchange due to large pure dephasing reduces the difference in population of states $|2\rangle$ and $|3\rangle$. Thus, the entanglement and all fidelities of the eigenstates becomes symmetric with respect to $J=0$. The fidelity of state $|\Psi_-\rangle$ is shown in Fig.~\ref{fig:OneFldTrpPDph}(c), and the fidelity of states $|1\rangle$ and $|2\rangle$ have a similar structure. In regions V, VI and VII, all eigenstates are evenly populated, such that the entanglement vanishes.

\begin{figure}[t]
 \centering
\includegraphics[width=0.9\columnwidth]{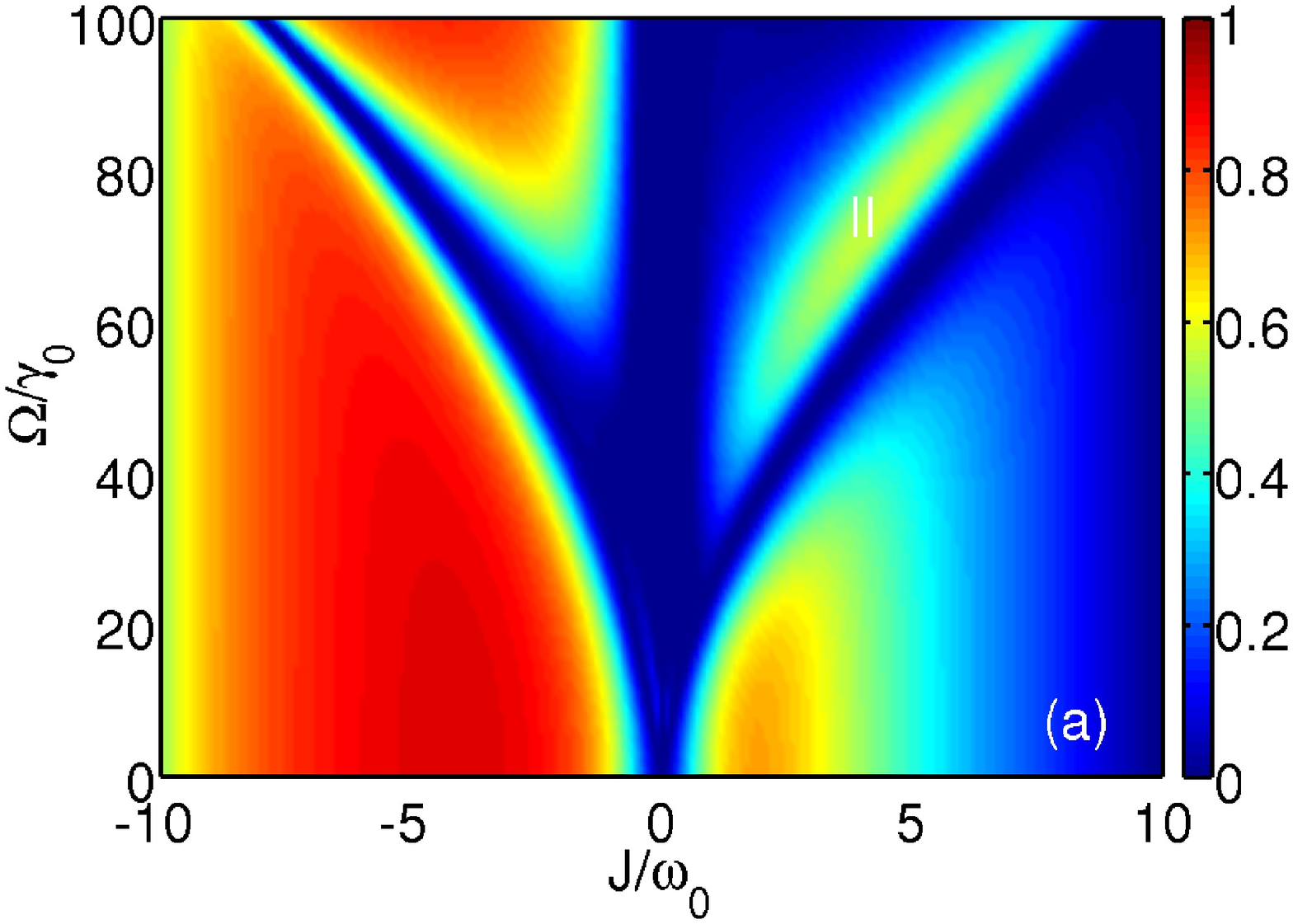}\\
\includegraphics[width=0.9\columnwidth]{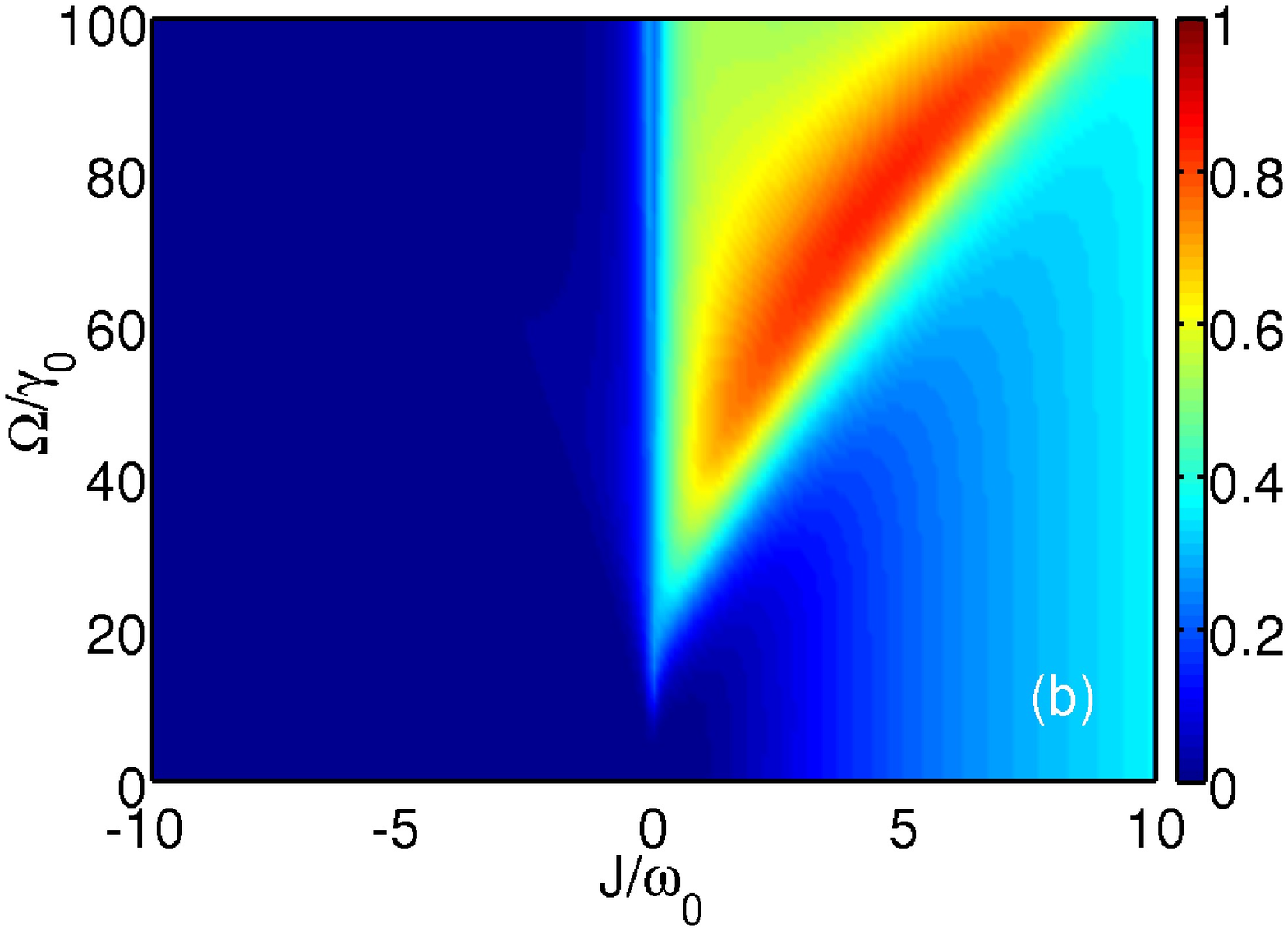}
\caption{\label{fig:LargRPDph}(Color online) Steady-state entanglement and interpretation in terms of Bell state populations against the driving field Rabi frequency $\Omega$ and the coupling $J$. The figure is similar to Fig.~\ref{fig:OneFldTrpPDph}, but generated for larger qubit separation $r_{21}=\lambda_0/50$. (a) shows the amount of entanglement,  and (b) the fidelity of the state $|\Psi_-\rangle$.}
\end{figure}

For larger separations such as $r_{21}=\lambda_0/50$, an entanglement of $0.58$ can be observed in region II of Fig.~\ref{fig:LargRPDph}(a). It arises mainly from contributions of the state $|\Psi_-\rangle$ because it is still created with a high fidelity of about $0.83$, see Fig.~\ref{fig:LargRPDph}(b).

\begin{figure}[t]
 \centering
\includegraphics[width=0.9\columnwidth]{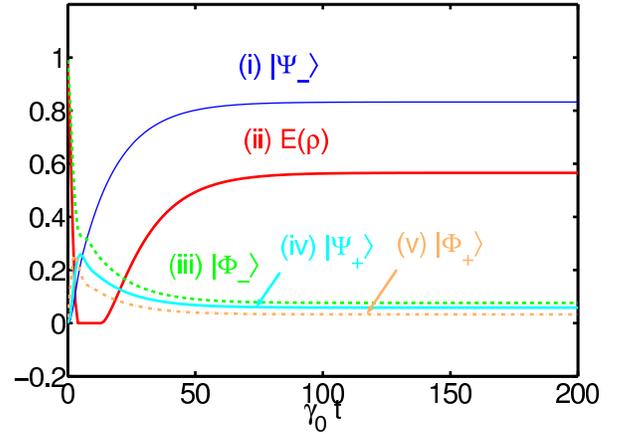}
\caption{\label{fig:OneFldt}(Color online) Time evolution of steady-state entanglement generation with a single cw field. The red curve shows (ii) the amount of entanglement, the blue thin line (i) shows the fidelity of state $|\Psi_-\rangle$,  the green dashed line (iii) that of $|\Phi_-\rangle$, the cyan line (iv) that of $|\Psi_+\rangle$, and the orange dot-dashed line (v) that of $|\Phi_+\rangle$. The parameters are $r_{21}=\lambda_0/50$, $p=1$, $\delta=2.5\gamma_0$, $\Gamma_\phi=0.01\gamma_0$ and $\Omega=80\gamma_0$.}
\end{figure}

We conclude by analyzing the time evolution of the system driven by one field for large separation $r_{21}=\lambda_0/50$, as shown in Fig.~\ref{fig:OneFldt}. Interestingly, the population first is pumped to the upper states $|2\rangle$ and $|1\rangle$, which decay to $|3\rangle$. From time $\gamma_0 t=3$ on, the damping of $|1\rangle$ is dominant. Therefore, the population of state $|3\rangle$ continually increases to is maximum value of $0.83$ at about $\gamma_0 t=80$. The system then remains in this steady state with $E(\rho)=0.58$.

\subsection{Trapping assisted by dark-state techniques}
Coherent population trapping (CPT) and the STIRAP technique are a convenient tool to prepare atoms in superposition states without decoherence throughout the preparation.  Typically, two driving fields are applied to an atom to generate a $\Lambda$-type level scheme. If the driving fields are chosen suitably, the system is driven into a dark state, which is a coherent superposition of the two ground states, with relative weights depending on the driving field parameters. Here, in contrast to previous implementations, we discuss coherent population trapping and adiabatic population transfer between two entangled states. Thereby, the entangled states are protected from decoherence, and the preparation is robust against noise in the field. 

\begin{figure}[t]
  \centering
  \includegraphics[width=0.9\columnwidth]{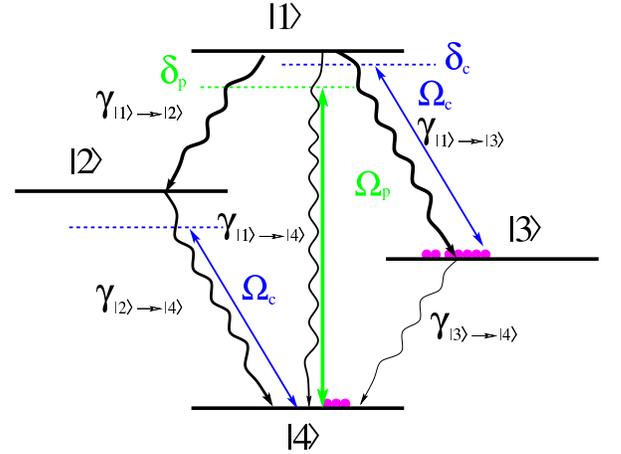}
  \caption{\label{fig:EITTrpSetup}(Color online) Level diagram and relevant coherent and incoherent processes for entanglement generation using dark state techniques.}
 \end{figure}

To enable dark state techniques, we apply two driving fields denoted as control field $\Omega_c$ and pumping field $\Omega_p$. These fields drive the two-photon and the $|1\rangle \leftrightarrow |3\rangle$ channels, respectively, as shown in Fig~\ref{fig:EITTrpSetup}. The transitions $|3\rangle \leftrightarrow |4\rangle$ and $|1\rangle \leftrightarrow |2\rangle$ are far off-resonant and can be neglected.
The control field has frequency $\omega_{L}^{(1)}=\omega_{L}^{(2)}=\omega_{c}$, the two-photon channel is driven with frequency $\omega_{L}^{(3)}=\omega_{p}$. We define a unitary transformation
\begin{subequations}
\begin{align}
           U &=\sum_j e^{i\omega_{Rj}t)|j\rangle \langle j|}  \,,\\
 \omega_{R1} &=\omega_4+\omega_p \,,\\
 \omega_{R2} &=\omega_4+\omega_c \,,\\
 \omega_{R3} &=\omega_4+\omega_p-\delta_c \,,\\
  \omega_{R4} &=\omega_4 \,.
\end{align}
\end{subequations}
The one-photon detuning is defined as $\delta_p=(\omega_1-\omega_4)-\omega_p$ and $\delta_c=(\omega_1-\omega_3)-\omega_c$. This gives
\begin{subequations}
 \begin{align}
 \Delta_1 &=\delta_p \,,\\
 \Delta_2 &=(\omega_2-\omega_4)-(\omega_1-\omega_3)+\delta_c \,,\\
 \Delta_3 &=\delta_p-\delta_c \,,\\
 \Delta_4 &=0 \,.
 \end{align}
\end{subequations}
In the RWA, the master equation for this scheme reads
\begin{equation}\label{eq:MEQEIT}
 \dot \rho=-i\left[\sum_j \Delta_j |j\rangle \langle j|, \rho \right]-\frac{i}{\hbar}[\bar{H}_I,\rho]+\tilde{\mathscr{L}}\rho+\mathscr{L}_{TP}\rho\,,
\end{equation}
where
\begin{align}\label{eq:EITHI}
\bar{H}_{I}&=\hbar \Omega_c\left[a\left(\beta+a p e^{-i\phi_c}\right)-b\left(\alpha+\beta p e^{-i\phi_c} \right)\right]R_{24} \,\nonumber \\\nonumber
&-\hbar \Omega_c \left[a\left(\beta-a pe^{-i\phi_c}\right)-b\left(\alpha-\beta p^{-i\phi_c}\right)\right]R_{13} \,\\
& +2\hbar \Omega_p (a^2-b^2)\left(1+e^{-i\phi_p}\right)R_{14}+{\textrm H.c.}\,.
\end{align}
Here, we have neglected the off-resonant excitations $|1\rangle \leftrightarrow |2\rangle$ and $|3\rangle \leftrightarrow |4\rangle$, as well as rapidly oscillating contributions from $H_{BIE}$. We also exclude rapidly oscillating parts relevant to $\gamma_{|2\rangle}^{cross}$ and $\gamma_{|3\rangle}^{cross}$ in $\mathscr{L}\rho$ yielding $\tilde{\mathscr{L}}\rho$. The phases are $\phi_{j}=2\pi (r_{21}/\lambda_0)(\omega_j/\omega_0)$ with $j\in \{c,p\}$. Unavoidable, the state $|2\rangle$ is coupled to the ground state by $\Omega_c$.

\begin{figure}[t]
 \centering
\includegraphics[width=0.9\columnwidth]{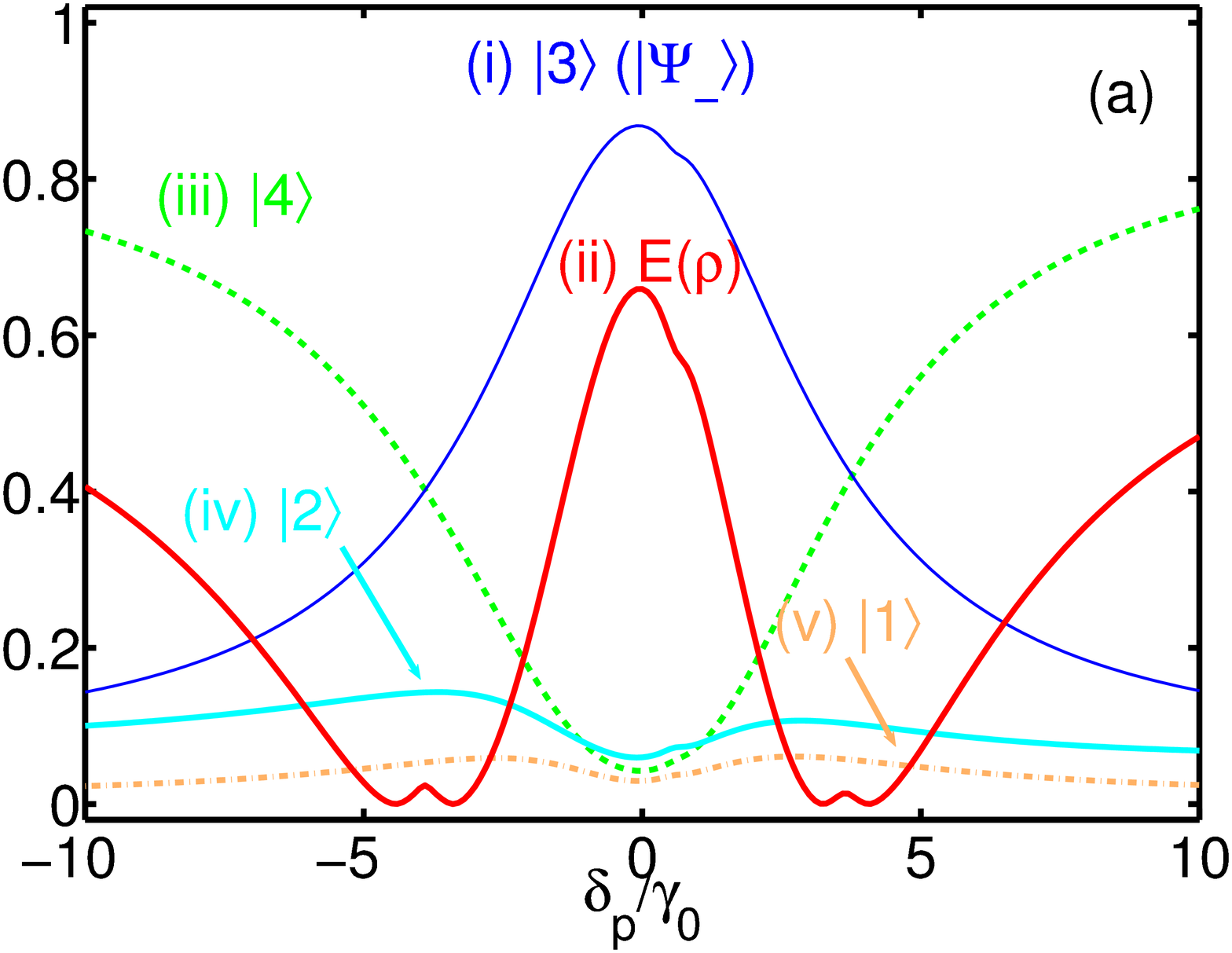}\\
\includegraphics[width=0.9\columnwidth]{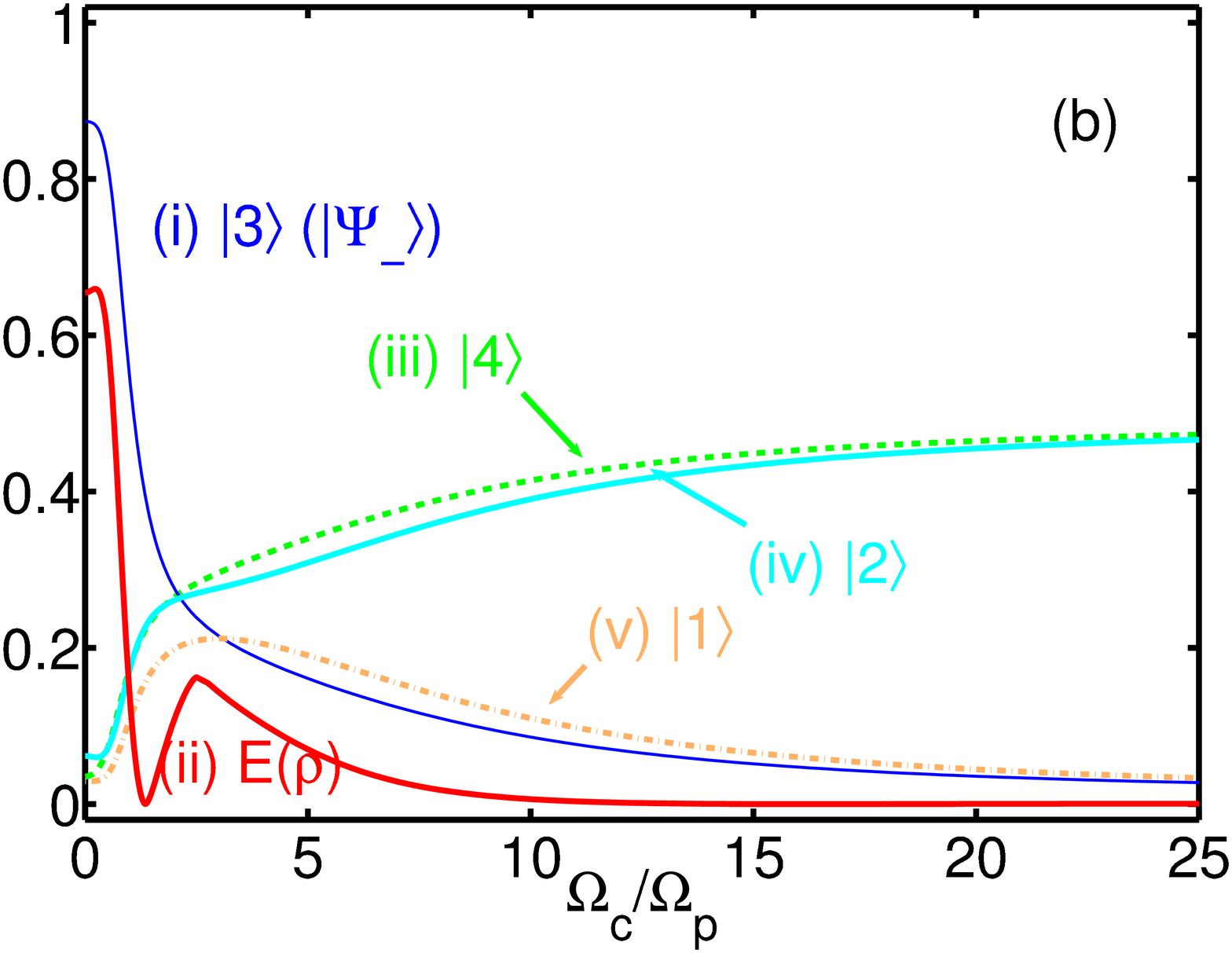}
\caption{\label{fig:EITTrp}(Color online) Entanglement generation with the dark state technique. In (a), the results are shown against the detuning $\delta_p$. The red curve (ii) shows the amount of entanglement, the blue thin curve (i) the fidelity of $|\Psi_-\rangle$, the green dashed line (iii) that of $|4\rangle$, the cyan line (iv) that if $|2\rangle$, and the orange dot-dashed line (v) that of $|1\rangle$. The parameters are $\delta_c=0$ and $\Omega_p=\gamma_0$.
Corresponding results shown against the control field Rabi frequency $\Omega_c$. The parameters are $\Omega_p=4\gamma_0$ and $\delta_p=\delta_c=0$. 
In both subpanels, the other parameters are $r_{21}=\lambda_0/50$, $\Gamma_\phi=0.01\gamma_0$, $\tilde{\gamma}_0=0.02\gamma_0$, $J=5\omega_0$ and $p=1$.}
\end{figure}

In Fig.~\ref{fig:EITTrp} (a), we demonstrate a coherent trapping of the intermediate state $|3\rangle$ via the dark state approach. This state  is of interest since it can approach the maximally entangled Bell states $|\Psi_\pm\rangle$, and the decay rate can be modified to be very small by the strong AOC.  The optimal trapping is obtained for $\delta_p=\delta_c=0$, $\Omega_p=4\gamma_0$, and $\Omega_c=\gamma_0$. Then, the fidelity of $|\Psi_-\rangle$ is about $0.87$ corresponding to an entanglement of $0.66$. Only about three percent population remains in the state $|1\rangle$. The intermediate state $|2\rangle$ is populated to about $6\%$.
Note that the state $|3\rangle$ could be either the symmetric state for $J>0$ or the antisymmetric state for a negative $J$.
The population in the state $|3\rangle$ can be controlled by the field $\Omega_c$, as shown in Fig.~\ref{fig:EITTrp}(b). For small $\Omega_c$, the state $|3\rangle$ is highly trapped, as the corresponding dark state has a high contribution of $|3\rangle$. For $\Omega_c\gg \Omega_p$, correspondingly the system is trapped in $|4\rangle$ and $|2\rangle$. When the pumping field $\Omega_p$ becomes comparable to the controlling field $\Omega_c$, the two entangled states are equally populated. The system is in a superposition state of two  maximally entangled states, which leads to vanishing overall entanglement as the superposition state is a separable state. As the controlling field increases further, the state $|2\rangle$ is excited. The population in this state is similar to that of the ground state. These two states together form a separable mixed state, and the system becomes disentangled.

We also investigated the time evolution in the dark-state approach. Similar to the trapping assisted by a single field, we found that it takes a rather long time ($>60\gamma_0^{-1}$) to reach the maximal value of entanglement and fidelity of the target state.

The STIRAP technique~\cite{RMP70p1003} is another robust method to coherently prepare a target state. In the following, we apply STIRAP to our system. For this, we use the same configuration shown in Fig.~\ref{fig:EITTrp} and sech-shaped pulses defined as 
\begin{subequations}
 \begin{align}
\label{eq:STIRAPPulses}
  \Omega_c &=3\Omega_0 \textrm{sech}[\pi (t-5\tau_c)/2\tau_{c}] \,,\\
  \Omega_p &=\Omega_0 \textrm{sech}[\pi (t-5\tau_p-\tau_d)/2\tau_{p}] \,,
 \end{align}
\end{subequations}
where $\tau_{i}$ with  $i\in \{c,p\}$ are the durations of the pulses, and $\tau_d$ the delay between the pulses. The factor $3$ is introduced to balance the prefactors of $R_{13}$ and $R_{14}$ in Eq.~(\ref{eq:EITHI}).
 
\begin{figure}[t]
 \centering
\includegraphics[width=0.9\columnwidth]{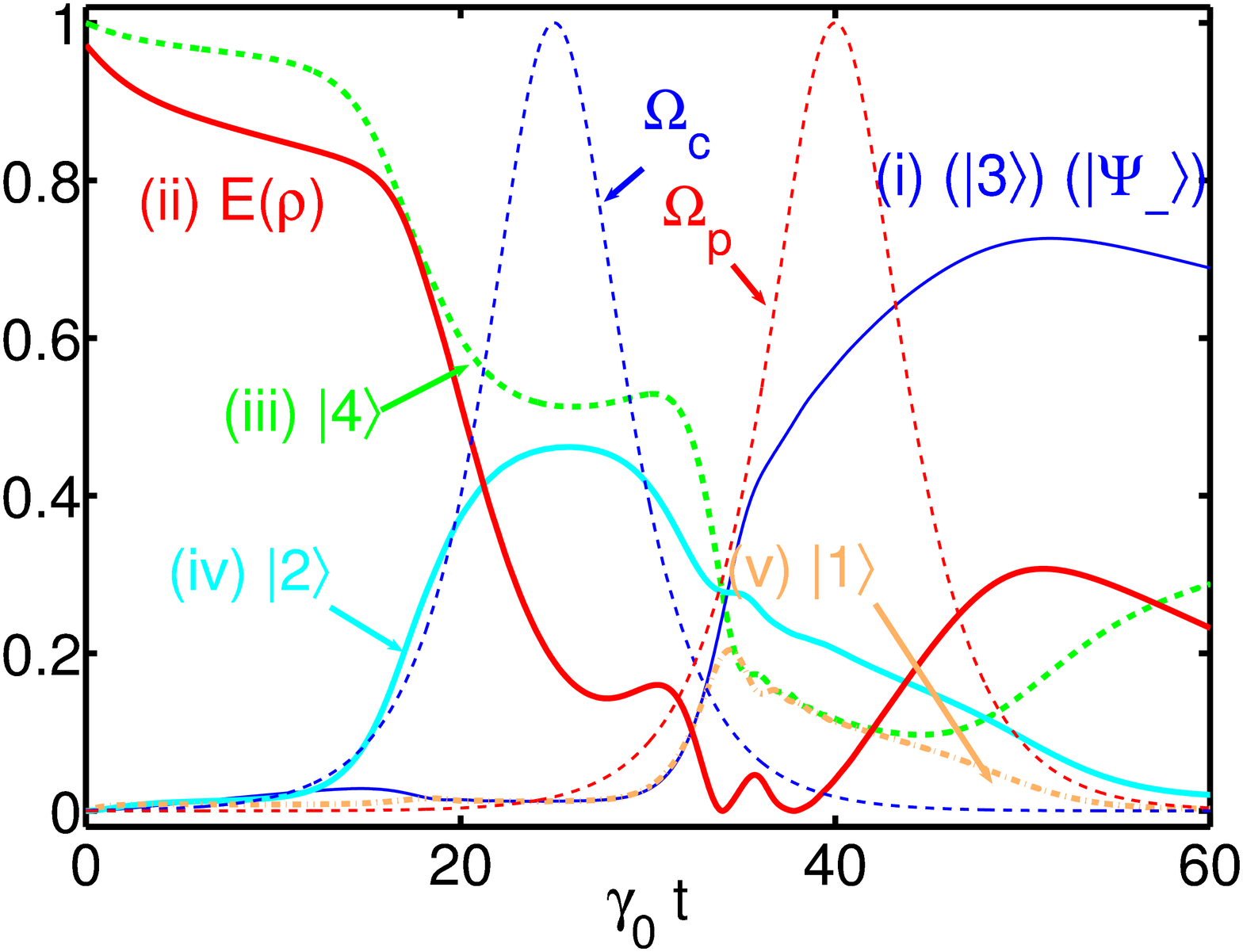}
\caption{\label{fig:STIRAP}(Color online) Coherent population transfer from the ground state to state $|3\rangle$ via the STIRAP technique. The Stokes pulse $\Omega_c$ (blue thin dashed line) precedes the pumping pulse $\Omega_p$ (red thin dashed line) by $15\gamma_0^{-1}$. In the figure, both pulses are shown normalized. 
The other curves show the amount of entanglement [red line (ii)], and the fidelity of states $|\Psi_-\rangle$ [blue thin line (i)], $|4\rangle$ [green dashed line (iii)],  $|2\rangle$ [cyan  line (iv)], and  $|1\rangle$ [orange dot-dashed line (v)].
The durations of both pulses are chosen as $\tau=5\gamma_0^{-1}$. The detunings are $\delta_p=\delta_c=\delta$. The other parameters are $J=5\omega_0$, $r_{21}=\lambda_0/50$, $\Gamma_\phi=0.01\gamma_0$, $\tilde{\gamma}_0=0.02\gamma_0$ and $p=1$.}
\end{figure}
We found that the CPT is optimal in two-photon resonance $\delta_p=\delta_c=0$ for $\tau_p=\tau_c=\tau$. As shown in Fig.~\ref{fig:STIRAP}, using parameters $\gamma_0 \tau=5$, $\tau_d=3\tau$ and $\Omega_0=10\gamma_0$, about $70\%$ population is transferred from the ground state $|\Phi_-\rangle$ to the state $|3\rangle$, which tends to $|\Psi_-\rangle$ for positive $J$ or $|\Psi_+\rangle$ if $J<0$.

\section{\label{summary}Summary}

In this work, we have derived and discussed the master equation for two interacting qubits  coupled to a common 1D bath. In particular, we have focused on the case of strong interaction between the qubits, with coupling rate exceeding the respective transition frequencies of the two qubits. Decoherence of the qubits is taken into account by coupling the system to a bath of harmonic oscillators. 
We have found that the strong coupling between the two qubits can lead to ground state entanglement even in the absence of any driving field, with the entangled state populated by spontaneous emission. We have also analyzed the case of a time-dependent strong qubit-qubit coupling. We found that starting from an initially unentangled ground state of the two qubits, a dynamical switching of the qubit coupling can lead to entanglement with high fidelity on time scales on the order of the inverse qubit transition frequency. 
Employing a single external driving field, both symmetric and antisymmetric collective qubit states can be prepared and maintained with  high fidelity, thus also leading to entanglement. Finally, using two external fields, we also demonstrated the possibility for entangled state preparation via a dark state approach or STIRAP technique in our configuration. Possible applications of our scheme include nonlocal quantum gates and quantum memories.

\begin{acknowledgments}
KX  gratefully acknowledges hospitality at Institute for Quantum Studies and Department of Physics and Astronomy, Texas A\&M University, College Station, Texas, USA, where finial reversion of this work was performed.
\end{acknowledgments}

\appendix

\section{\label{app-decay}Decay rates}
In this appendix we derive the cross decay rate and the decay rate of isolated qubits. We model the bath as an open 1D transmission line~\cite{Science327p840} with total inductance $L_r=Ll$ and capacitance $C_r=Lc$, where $l$ and $c$ are the inductance and capacitance per unit length, respectively. $L$ is the length of the line. The zero point fluctuations of the current in the transmission line is $I_r=\sqrt{\hbar \omega_r/2Ll}$. The coupling $\eta_k^{(l)}$ introduced in  Eq.~(\ref{eq:HB}) reads~\cite{PRL96p127006}
\begin{align}
\hbar\eta_k^{(l)}=\vec{e}_k \bullet \vec{d}^{(l)} M_B^{(l)}I_p^{(l)}I_r\,.
\end{align}

The single photon interaction Hamiltonian between the qubits and the bath takes the form
\begin{equation}
 H_{BSP}=\hbar \sum_{l=1}^2 \sigma_x^{(l)}\sum_k \eta_k^{(l)} \left(e^{kr_l}a_k+H.c.\right)\,,
\end{equation}
with polarization $\vec{e}_k$ of the current, which only takes two directions.  $r^{(l)}$ indicates the position of the $l$th qubit in the vacuum field. The wave number of the $k$th mode is $k=\omega_r/v$, with the wave phase velocity $v=1/\sqrt{lc}$.

According to Eq.~(\ref{eq:HB}), the cross decay rate is given by
\begin{equation}
 \gamma_{21}=\sum_k \pi \delta(\omega-\in) |\eta_k|^2 \left(\vec{e}_k \bullet \vec{d}\right)^2 \cos(kr_{21})\,,
\end{equation}
where $r_{21}=r_2-r_1$. We consider the transmission line as an Ohmic bath with a high cutoff frequency $\omega_{cut}$ such that $\sum_k \left(\vec{e}_k \bullet \vec{d}\right)^2 \rightarrow 2\frac{L}{2\pi v}\int_0^\infty e^{-\omega/\omega_{cut}} d\omega$. Here we have introduced an exponential cutoff \cite{OhmicBath}. Then, the rate $\gamma_{12}$ can be evaluated as
\begin{align}
 \gamma_{21}&=\frac{L}{\pi v} \sin^2\theta \frac{\left(M_BI_p\right)^2}{\hbar^2}\int_0^\infty \pi \delta(\omega-\in)\frac{\hbar \omega}{2L_r}\cos(kr_{21}) e^{-\omega/\omega_{cut}} d\omega \nonumber\\
 &=\sin^2\theta \frac{\left(M_BI_p\right)^2}{2\hbar z}\in \cos\left(\frac{\in}{v}r_{21}\right)\nonumber\\
 &=\gamma_0(\in)\cos\left(\frac{\in}{v}r_{21}\right) \,,
\end{align}
where $\in$ is the transition frequency between the two involved states. For simplicity, we have assumed $\vec{d}_1=\vec{d}_1=\vec{d}$. Here we define the decay rate for the isolated qubit as
\begin{align}
\gamma_0(\omega)=\sin^2\theta \frac{\left(M_BI_p\right)^2}{2\hbar z} \omega\,.
\end{align}
Similarly, we derived the two-photon decay rates $\tilde{\gamma}_0$ and $\tilde{\gamma}_{12}$.
Following the diagonalization, the rates in Eq.~(\ref{eq:Lrho}) are given by
\begin{subequations}
\begin{align}
\gamma_{|1\rangle\rightarrow |2\rangle}&=2(1+4ab\alpha\beta)\left(\gamma_0^{(-)}+2\alpha \beta p \gamma_{12}^{(-)}\right)\,,\\
\gamma_{|2\rangle\rightarrow |4\rangle}&=2(1-4ab\alpha\beta) \left(\gamma_0^{(+)}+2\alpha \beta p \gamma_{12}^{(+)}\right)\,,\\
\gamma^{(cross)}_{|2\rangle}&=(a^2-b^2)\left[2\alpha\beta(\gamma_0^{(-)}+\gamma_0^{(+)})+p(\gamma_{12}^{(-)}+\gamma_{12}^{(+)})\right]\,,\\
\gamma_{|1\rangle\rightarrow |3\rangle}&=2(1-4ab\alpha\beta)\left(\gamma_0^{(+)}-2\alpha \beta p \gamma_{12}^{(+)}\right)\,,\\
\gamma_{|3\rangle\rightarrow |4\rangle}&=2(1+4ab\alpha\beta)\left(\gamma_0^{(-)}-2\alpha \beta p \gamma_{12}^{(-)}\right)\,,\\
\gamma^{(cross)}_{|3\rangle}&=(a^2-b^2)\left[2\alpha\beta(\gamma_0^{(-)}+\gamma_0^{(+)})-p(\gamma_{12}^{(-)}+\gamma_{12}^{(+)})\right]\,,
\end{align}
\end{subequations}
where the dipole-dipole related contributions are listed in Appendix~\ref{app-shift}.

\section{\label{app-shift}Dipole-dipole induced couplings}
The coupling coefficients related to the dipole-dipole interaction can be evaluated to give
\begin{subequations}
\begin{align}
\Omega_{(d,\pm)}&=-\left(\Omega_{(d,\pm)}^{+} + \Omega_{(d,\pm)}^{-}\right)\nonumber\\
&= \frac{\gamma_0^{(\pm)}}{\pi} \mathfrak{Re}\left[e^{-S_\pm}\Gamma(0,-S_\pm)-e^{S_\pm}\Gamma(0,S_\pm) \right . \nonumber\\
& \quad \left. -i\pi e^{S_\pm} +\frac{2}{S_\pm} \right]\,,\\
\Omega_{(d,\pm)}^+&= \frac{\gamma_0(\bar w\pm |J|)}{\pi (\bar w\pm |J|)} P \int_0^\infty \frac{\omega}{\omega+(\bar w\pm |J|)} \nonumber\\
&\quad \times \cos\left (\frac{\omega}{v}r_{21}\right )e^{\frac{\omega}{\omega_{cut}}}d\omega  \nonumber \\
&=-\frac{\gamma_0^{(\pm)}}{\pi} \mathfrak{Re}\left[e^{-S_\pm}\Gamma (0,-S_\pm)+\frac{1}{S_\pm} \right] \,,\\
\Omega_{(d,\pm)}^-&=\frac{\gamma_0(\bar w\pm |J|)}{\pi (\bar w\pm |J|)} P \int_0^\infty \frac{\omega}{\omega-(\bar w\pm |J|)}\nonumber\\
&\quad  \times \cos\left (\frac{\omega}{v}r_{21}\right)e^{\frac{\omega}{\omega_{cut}}}d\omega \nonumber\\ 
&=\frac{\gamma_0^{(\pm)}}{\pi}\mathfrak{Re}\left[e^{S_\pm}\Gamma (0,S_\pm)+\pi i e^{S_\pm}-\frac{1}{S_\pm} \right] \,,\\
\tilde\Omega_d^{(+)}&=-\frac{\tilde\gamma_0}{\pi} \mathfrak{Re}\left[e^{-\tilde{S}}\Gamma (0,-\tilde{S})+\frac{1}{\tilde{S}} \right] \,,\\
\tilde\Omega_d^{(-)}&=-\frac{\tilde\gamma_0}{\pi} \mathfrak{Re} \left[e^{\tilde{S}}\Gamma (0,\tilde{S})+\pi i e^{\tilde{S}}-\frac{1}{\tilde{S}} \right]\,,
\end{align}
\end{subequations}
where
\begin{subequations}
\begin{align}
 \tilde{S} &=\frac{2\bar w}{\omega_0}\left[-\frac{1}{\varsigma_c}+i2\pi \frac{r_{21}}{\lambda_0}\right]\,,\\
 S_\pm &=\frac{\bar w\pm |J|}{\omega_0}\left[-\frac{1}{\varsigma_c}+i2\pi \frac{r_{21}}{\lambda_0}\right]\,,
\end{align}
\end{subequations}
with $\varsigma_c=\omega_c/\omega_0$, and $\omega_c$ is the exponential cutoff frequency of the Ohmic bath. $\Gamma$ is the incomplete gamma function. Furthermore, 
\begin{subequations}
\label{eq:Rlx}
\begin{align}
 \gamma_0^{(\pm)}&=\frac{\bar w\pm |J|}{\omega_0}\gamma_0(\omega_0)\,,\\
\gamma_{12}^{(\pm)}&=\gamma_0^{(\pm)} \cos\left(\frac{\bar w\pm |J|}{\omega_0}2\pi r_{21}/\lambda_0\right)\,,\\
 \tilde \gamma_0&=\frac{2\bar w}{\omega_0}\tilde\gamma_0(\omega_0)\,, \\
\tilde \gamma_{12}&=\tilde \gamma_0 \cos\left(\frac{2\bar w}{\omega_0} 2\pi r_{21}/\lambda_0\right)\,.
\end{align}
\end{subequations}

\section{\label{app-dd}Dipole-Dipole shifts and bath-induced excitations}
The coherent part $H_{DD}$ includes the dipole-dipole shifts $H_{DDS}=\hbar \sum_j\omega_j^{(DDS)}|j\rangle \langle j|$ in the transition frequencies, where
%
\begin{subequations}\label{eq:frq}
 \begin{align}
  \omega^{(DDS)}_1&\sim -2(\alpha \beta+ab)p \Omega^-_{(d,-)}+2(\alpha \beta-ab)p \Omega^-_{(d,+)}\nonumber\\
 &\quad -2(a^2-b^2)^2\tilde\Omega^-_{d}\,,\\
  \omega^{(DDS)}_2 &\sim -2(\alpha \beta+ab)p \Omega^+_{(d,-)}-2(\alpha \beta-ab)p \Omega^-_{(d,+)}\,,\\
  \omega^{(DDS)}_3&\sim 2(\alpha \beta-ab)p \Omega^+_{(d,+)}+2(\alpha \beta+ab)p \Omega^-_{(d,-)}\,,\\
  \omega^{(DDS)}_4&\sim -2(\alpha \beta-ab)p \Omega^+_{(d,+)}+2(\alpha \beta+ab)p \Omega^+_{(d,-)} \nonumber \\
 & -2(a^2-b^2)^2\tilde\Omega^+_{d}\,.
 \end{align}
\end{subequations}
%
The bath-induced excitation $H_{BIE}$ is given by
\begin{equation}
 H_{BIE}=H_{BIE}^{(1)}+H_{BIE}^{(2)}\,,
\end{equation}
where
%
 \begin{align}\nonumber
H_{BIE}^{(1)}= &-i (a^2-b^2) p  \left(\Omega^+_{(d,+)}-\Omega_{(d,-)}^+\right)(R_{24}\rho R_{21}-R_{12}\rho R_{42})\,,\\\nonumber
&-i (a^2-b^2) p\left(\Omega_{(d,+)}^--\Omega_{(d,-)}^-\right)(R_{42}\rho R_{12}-R_{21}\rho R_{24})\,,\\\nonumber
H_{BIE}^{(2)}= &-i (a^2-b^2)p  \left(\Omega_{(d,-)}^+-\Omega_{(d,+)}^+\right)(R_{34}\rho R_{31}-R_{13}\rho R_{43})\,,\\\nonumber
&-i (a^2-b^2)p \left(\Omega_{(d,-)}^--\Omega_{(d,+)}^-\right)(R_{43}\rho R_{13}-R_{31}\rho R_{34})\,.
 \end{align}
$H_{BIE}$ is negligible because it arises from small dipole-dipole shifts and is suppressed by a factor of $a^2-b^2$.

\section {\label{app-basis}Decay and pumping in the collective basis}
Here we list the decay, pumping and coherent interaction contributions in the collective state basis. They are derived from the dissipation in the dressed-state basis.

The relaxation in terms of the collective state basis becomes 
\begin{subequations}
\begin{align}
\mathscr{D}_{1}\rho&= a^2 \gamma_{|1\rangle\rightarrow |2\rangle} \left\{ R_{2E}\rho R_{E2}-\frac{1}{2}R_{EE}\rho -\frac{1}{2}\rho R_{EE} \right\}\,\nonumber\\
&+ a^2 \gamma_{|2\rangle\rightarrow |4\rangle} \left\{ R_{G2}\rho R_{2G}-\frac{1}{2}R_{22}\rho -\frac{1}{2}\rho R_{22} \right\}\,\nonumber\\
&+a^2 \gamma^{(cross)}_{|2\rangle} \left\{R_{2E}\rho R_{2G}+R_{G2}\rho R_{E2}\right\}\,,\\
\mathscr{D}_{2}\rho&= a^2 \gamma_{|1\rangle\rightarrow |3\rangle} \left\{ R_{3E}\rho R_{E3}-\frac{1}{2}R_{EE}\rho-\frac{1}{2}\rho R_{EE} \right\}\,\nonumber\\
&+ a^2 \gamma_{|3\rangle\rightarrow |4\rangle}
\left\{ R_{G3}\rho R_{3G}-\frac{1}{2}R_{33}\rho- \frac{1}{2}\rho R_{33}\right\}\,\nonumber\\
&-a^2 \gamma^{(cross)}_{|3\rangle}\left\{R_{G3}\rho R_{E3}+R_{3E}\rho R_{3G}\right\}\,,\\
\mathscr{D}_{TP}\rho&= a^2 \gamma_{TP} \left\{ R_{GE}\rho R_{EG}-\frac{1}{2}R_{EE}\rho-\frac{1}{2}\rho R_{EE} \right\}\,\nonumber\\
&+ ab \gamma_{TP}\left\{R_{GG}\rho R_{GG}-\frac{1}{2}R_{GE}\rho-\frac{1}{2} \rho R_{EG} \right\}\,.
\end{align}
\end{subequations}

The pumping contributions are given by
\begin{subequations}
\begin{align}
\mathscr{P}_{1}\rho&= b^2 \gamma_{|1\rangle\rightarrow |2\rangle} \left\{ R_{2G}\rho R_{G2}-\frac{1}{2}R_{GG}\rho -\frac{1}{2}\rho R_{GG} \right\}\,\nonumber\\
&+ b^2 \gamma_{|2\rangle\rightarrow |4\rangle} \left\{ R_{E2}\rho R_{2E}-\frac{1}{2}R_{22}\rho -\frac{1}{2}\rho R_{22} \right\}\,\nonumber\\
&-b^2 \gamma^{(cross)}_{|2\rangle} \left\{R_{E2}\rho R_{G2}+R_{2G}\rho R_{2E}\right\}\,,\\
\mathscr{P}_{2}\rho&= b^2 \gamma_{|1\rangle\rightarrow |3\rangle} \left\{ R_{3G}\rho R_{G3}-\frac{1}{2}R_{GG}\rho-\frac{1}{2}\rho R_{GG} \right\}\,\nonumber\\
&+ b^2 \gamma_{|3\rangle\rightarrow |4\rangle}
\left\{ R_{E3}\rho R_{3E}-\frac{1}{2}R_{33}\rho- \frac{1}{2}\rho R_{33}\right\}\,\nonumber\\
&+b^2 \gamma^{(cross)}_{|3\rangle} \left\{R_{E3}\rho R_{G3}+R_{3G}\rho R_{3E}\right\}\,,\\
 \mathscr{P}_{TP}\rho&= b^2 \gamma_{TP} \left\{ R_{EG}\rho R_{GE}-\frac{1}{2}R_{GG}\rho-\frac{1}{2}\rho R_{GG} \right\} \,\nonumber\\
&+ ab \gamma_{TP} \left\{R_{EE}\rho R_{EE}-\frac{1}{2}R_{EG}\rho-\frac{1}{2} \rho R_{GE} \right\}\,.
\end{align}
\end{subequations}
The dissipation in the dressed state representation introduces a coherent contribution in the collective basis as
\begin{subequations}
 \begin{align}
  \mathscr{C}_{1}\rho=&ab \gamma_{|1\rangle\rightarrow |2\rangle} \left\{ R_{2E}\rho R_{G2}-\frac{1}{2}R_{GE}\rho -\frac{1}{2}\rho R_{GE} \right\}\,\nonumber\\
  &+ ab \gamma_{|1\rangle\rightarrow |2\rangle} \left\{ R_{2G}\rho R_{E2}-\frac{1}{2}R_{EG}\rho -\frac{1}{2}\rho R_{EG} \right\}\,\nonumber\\
  &-ab \gamma_{|2\rangle\rightarrow |4\rangle}  \left\{ R_{G2}\rho R_{2E}+R_{E2}\rho R_{2G} \right\}\,\nonumber\\
  &-ab \gamma^{(cross)}_{|2\rangle} \left\{R_{E2}\rho R_{E2}+R_{2E}\rho R_{2E}\right\}\,\nonumber\\
&+ab \gamma^{(cross)}_{|2\rangle}
\left\{R_{G2}\rho R_{G2}+R_{2G}\rho R_{2G}\right\}\,,\\
\mathscr{C}_{2}\rho=&ab \gamma_{|1\rangle\rightarrow |3\rangle} \left\{R_{3E}\rho R_{G3}-\frac{1}{2}R_{GE}\rho -\frac{1}{2}\rho R_{GE}\right\} \,\nonumber\\
&+ab \gamma_{|1\rangle\rightarrow |3\rangle} \left\{R_{3G}\rho R_{E3}-\frac{1}{2}R_{EG}\rho -\frac{1}{2}\rho R_{EG}\right\} \,\nonumber\\
&-ab \gamma_{|3\rangle\rightarrow |4\rangle} \left\{R_{E3}\rho R_{3G}+R_{G3}\rho R_{3E} \right\} \,\nonumber\\
&-ab \gamma^{(cross)}_{|3\rangle} \left\{R_{G3}\rho R_{G3}+R_{3G}\rho R_{3G} \right\} \nonumber\,\\
&+ab \gamma^{(cross)}_{|3\rangle} \left\{R_{E3}\rho R_{E3}+R_{3E}\rho R_{3E} \right\} \,.
 \end{align}
\end{subequations}
Note that the relation $a^2>b^2$ holds. Thus, the decay rates are always larger than the corresponding pumping rates.

%

\end{document}